\documentclass[draft]{agujournal2019}
\usepackage{url} 
\usepackage[inline]{trackchanges} 
\usepackage{soul}
\usepackage{amsmath}

\newcommand{\CO}{CO$_2$}
\newcommand{\HHO}{H$_2$O}
\newcommand{\mum}{$\mu$m}
\newcommand{\Ls}{L$_s$}

\newcommand{\sigO}{1-$\sigma$}
\newcommand{\tiu}{J~m$^{-2}$~K$^{-1}$~s$^{1/2}$}
\newcommand{\tco}{T$_{\rm{CO}_2}$}

\draftfalse

\journalname{JGR Planets}

\begin{document}

%
%


\title{Observations of Water Frost on Mars with THEMIS: Application to the Presence of Brines and the Stability of (Sub)Surface Water Ice}

%
%




\authors{L.Lange\affil{1}, S.Piqueux\affil{2},  C.S.Edwards\affil{3}, F.Forget\affil{1}, J.Naar\affil{4}, E.Vos\affil{1}, A.Szantai\affil{1}}

\affiliation{1}{Laboratoire de Météorologie Dynamique, Institut Pierre-Simon Laplace (LMD/IPSL), Sorbonne Université, Centre National de la Recherche Scientifique (CNRS), École Polytechnique, École Normale Supérieure (ENS), Paris, France}
\affiliation{2}{Jet Propulsion Laboratory, California Institute of Technology, Pasadena, CA 91109, USA}
\affiliation{3}{Northern Arizona University, Department of Astronomy and Planetary Science PO BOX 6010 Flagstaff, AZ 86011, USA}
\affiliation{4}{Laboratoire Atmosphères, Milieux, Observations Spatiales (LATMOS/CNRS), Paris, France}

\correspondingauthor{Lucas Lange}{lucas.lange@lmd.ipsl.fr}



\begin{keypoints}
 \item Seasonal water frosts on Mars are identified with coincident visible and temperature data obtained with THEMIS poleward of 48\textdegree N and 21\textdegree S;
\item Water frosts remain too cold to melt as pure ice. However, the warmest frost deposits observed may co-exist with brines; 
 \item Warm water frost, which releases a large amount of water vapor when it sublimates, cannot stabilize the low-latitude subsurface ice.
\end{keypoints}

%
%

%
%

\begin{abstract}
Characterizing the exchange of water between the Martian atmosphere and the (sub)surface is a major challenge for understanding the mechanisms that regulate the water cycle. Here we present a new dataset of water ice detected on the Martian surface with the Thermal Emission Imaging System (THEMIS). The detection is based on the correlation between bright blue-white patterns in visible images and a temperature measured in the infrared that is too warm to be associated with \CO~ice and interpreted instead as water ice. Using this method, we detect ice down to 21.4\textdegree S, 48.4\textdegree N, on pole-facing slopes at mid-latitudes, and on any surface orientation poleward of 45\textdegree~latitude. Water ice observed with THEMIS is most likely seasonal rather than diurnal. Our dataset is consistent with near-infrared frost detections and predictions by the Mars Planetary Climate Model. Water frost average temperature is 170~K, and the maximum temperature measured is 243~K, lower than the water ice melting point. Melting of pure water ice on the surface is unlikely due to cooling by latent heat during its sublimation. However, 243 THEMIS images show frosts that are hot enough to form brines if salts are present on the surface. The water vapor pressure at the surface, calculated from the ice temperature, indicates a dry atmosphere in early spring, during the recession of the CO$_2$ ice cap. The large amount of water vapor released by the sublimation of warm frost cannot stabilize subsurface ice at mid-latitudes.

\end{abstract}

\section*{Plain Language Summary}
During spring, parts of the Martian surface at mid and high latitudes are covered by a thin, bright layer of frost. Some of these ice deposits are made of CO$_2$ ice, but some frosts have a temperature that is too warm to be  CO$_2$ frosts and are thus constituted of water ice that forms in winter and sublimes in spring. We conducted a spatial and temporal mapping of these ice deposits using the camera THEMIS onboard the Mars Odyssey orbiter to better characterize the exchange of volatiles between the Martian atmosphere and the surface. Water ice is preferentially detected on pole-facing slopes rather than flat surfaces below 45\textdegree~latitude and all types of surfaces at higher latitudes. In late spring, water ice cannot melt despite solar heating because it cools down with the release of latent heat. On the other hand, these ice deposits are warm enough to form brines if salt crystals are present at the surface. When the frost sublimes, it serves as a source of water vapor that can diffuse into the ground and recondense to stabilize the permafrost below the surface. However, this effect is not sufficient to stabilize subsurface ice at low latitudes.

\section{Introduction}

Although water is a minor component of the Martian atmosphere, the water cycle is one of the most important contributors to the present and past climate of Mars. For example, although the current atmospheric moisture is low \cite<$\sim$13 microns precipitable,>{Smith2002}, water clouds have a strong influence on the atmospheric dynamics \cite{Wilson2007, Wilson2008, Madeleine2012, Navarro2014}. Similarly, the presence of ice buried in the subsurface has a significant impact on the surface energy balance, delaying or even preventing \CO~condensation in winter \cite{Haberle2008}. Therefore, characterizing the mechanisms controlling the water cycle is crucial to better understand the climate on Mars. This water cycle is mainly driven by the sublimation cycle of massive water-ice deposits located primarily at the North Pole and to a lesser extent at the South Pole, although the role of the regolith in this cycle remains controversial \cite<see a full discussion in>{Montmessin2017}. Every summer, the northern polar water ice cap sublimes, releasing water vapor that is then transported to lower latitudes. Depending on temperature and humidity conditions, this water vapor can condense in the atmosphere and form clouds \cite<e.g.,>{Curran1973}, diffuse into the subsurface \cite<e.g.,>{Schorghofer2005}, or form frost on the surface \cite<e.g.,>{Svitek1990}.

The characterization of surface water frost deposits is of great interest to assess 1) the exchanges between perennial water ice reservoirs at the poles and the rest of the planet \cite{BAPST2015}, 2) their potential to form liquid water or brines on the surface \cite<e.g.,>{Schorghofer2020}, 3) their contribution to current surface processes \cite<e.g.,>{Diniega2021, Dundas2021}, and 4) the exchanges of water vapor between surface frost and subsurface water ice buried in the mid and high latitudes \cite<e.g.,>{BAPST2015, Williams2015, Lange2023ice}. Since \citeA{Leighton1966}'s pioneering work on volatile exchanges between reservoirs, local approaches at lander landing sites  \cite{Svitek1990, Landis2007, MARTINEZ2016}, or globally from orbit  \cite{Kieffer2001, Titus2003, Schorghofer2006, Landis2007, Piqueux2008, Carrozzo2009, Schmidt2009, Vincendon2010water, Appr2011, Kereszturi2011, BAPST2015, Vincendon2015, Stcherbinine2023} have been undertaken to map and characterize water frost deposits. For the latter approach, two methods exist to date: 1) spectroscopic detections of water ice based on near-infrared spectral absorptions \cite{Langevin2007, Schmidt2009, Vincendon2010water, Kereszturi2011, Vincendon2015} and 2) the identification of bright patches with a temperature that is too warm to be CO$_2$ ice but is instead H$_2$O ice \cite{Kieffer2001, Titus2003, Schorghofer2006, Piqueux2008, BAPST2015, Stcherbinine2023}. This last technique requires simultaneous measurement in visible wavelengths, to detect frost through its distinct albedo or color, and at infrared wavelengths, to determine the composition of the ice through its temperature. This last quantity is required to determine the composition of the ice detected via a bright pattern in the visible range.  For instance, \citeA{Schorghofer2006} identified bright white patches on pole-facing slopes at low latitudes. While their thermal models suggest that this ice was indeed CO$_2$ ice, they were only able to definitively show that some bright patterns were indeed CO$_2$ frost via temperature measurements of these slopes.  As shown by  \citeA{Vincendon2014} and \citeA{Vincendon2015}, although some bright deposits are located on slopes where CO$_2$ ice is predicted by the models, spectroscopic measurements show that they can sometimes actually be composed of water ice instead of CO$_2$ ice. Thus, the observation of bright deposits on slopes and correlation with model prediction is not sufficient to demonstrate with certainty the nature of the ices observed.

The coincident acquisition of visible and temperature data enables us to distinguish CO$_2$ versus H$_2$O ices, and to document new properties for Mars such as:

\begin{enumerate}
    \item The temperature of water ice. This measurement can be used to constrain the thermo-physical properties of ice through its diurnal/seasonal evolution \cite<e.g.,>{Bapst2019},  or to determine whether ice can reach melting temperature. However, most of these measurements have been acquired on the massive, perennial water ice deposits at the poles, where temperatures barely exceed 200~K \cite{Kieffer2001, Titus2003, Piqueux2008, Bapst2019}. \citeA{BAPST2015} have measured ice temperatures down to $\pm$45\textdegree~latitude with the Thermal Emission Spectrometer \cite<TES, >{Christensen2001} and found higher water frost temperature ($\sim$220~-~240~K). However, they acknowledged that these warm-water ice temperatures might be due to sub-pixel terrain mixing (TES has a resolution of $\sim$3~×~6 km) and uncertainty in the retrieved surface temperatures \cite{Bapst2019}. \citeA{Carrozzo2009}, through a spectroscopic study, have detected water ice at tropical latitudes, which exhibits high temperatures (most between 180~K and 245~K, up to 260~K for few detections). \citeA{Vincendon2010water} and \citeA{Vincendon2015}  have extended their study, also detecting water frost on pole-facing crater slopes at tropical latitudes,  but they did not retrieve the temperatures of these deposits.
    
    \item The near-surface water vapor content. This quantity, while crucial for the stability of subsurface ice \cite{Schorghofer2005} or to constrain the vertical profile of water vapor in the lower atmosphere \cite<>[and references therein]{Leung2024}, has not been directly measured by any rover or lander to date. It has been indirectly derived from simultaneous measurements of relative humidity, temperature, and pressure measurements at the Phoenix \cite{Zent2010, Fischer2019}, Mars Science Laboratory \cite{Harri2014}, and Mars~2020 \cite{Polkko2023} landing sites. However, if water frost is present at the surface, the water vapor content at the surface can be directly constrained as the vapor pressure is equal to the vapor pressure over ice at saturation $p_{\rm{sat, ice}}$~(Pa),  which is a function of the ice temperature $T_{\rm{ice}}$~(K) \cite{Murphy2005}:
        \begin{equation}
        p_{\rm{sat, ice}} = \exp\left( -\frac{6143.7}{T_{\rm{ice}}} + 28.9074 \right)
        \label{eq:murphypsv}
    \end{equation}
    Therefore, determining water ice temperatures could provide a new, broader set of near-surface water vapor data. Yet, this analysis has not been conducted by studies measuring water ice temperatures \cite<e.g.,>{Kieffer2001, Titus2003, Piqueux2008, BAPST2015, Bapst2019}.
\end{enumerate}

Here we extend the previous studies on the presence of water frost on the Martian surface by presenting a new dataset of water ice observed at mid and high latitudes and by providing a unique dataset of water ice temperature and vapor pressure using measurements from the Thermal Emission Imaging System \cite<THEMIS, >{Christensen2004} onboard Mars Odyssey. The latter simultaneously combines visible and infrared measurements, enabling the detection and characterization of ice as performed by \citeA{Titus2003, Piqueux2008}, and \citeA{Wagstaff2008}  for water ice;  \citeA{Khuller2021frost} and \citeA{Lange2022a} for \CO~ice. \citeA{Titus2003, Piqueux2008} and \citeA{Wagstaff2008} focused their studies on the detection of water ice at high latitudes (at the South Pole and above 60\textdegree N respectively). Here we propose to extend their approach to the entire planet. 

High-resolution imagery allows the detection of small deposits of water ice found at low latitudes \cite{Carrozzo2009, Vincendon2010water, Vincendon2015}, previously inaccessible to \citeA{Kieffer2001} and \citeA{BAPST2015} studies, and reduces uncertainty induced by sub-grid mixing. In addition, measuring ice temperatures reveals whether ice can melt or form brines and constrain the near-surface water content. These data can also validate water frost predictions made by the Mars Planetary Climate Model \cite<PCM,>{Forget1999, Lange2023Model}. This model, along with the processing of THEMIS data, is presented in section \ref{sec:methods}. The results on water frost spatio-temporal distribution, temperatures, and derived near-surface water vapor are described in section  \ref{sec:results}.  Section \ref{sec:discussion}  focuses on the discussion of these results, and the conclusions are summarized in section\ref{sec:conclusions}. 

\section{Methods \label{sec:methods}}
\subsection{THEMIS Dataset}

This study uses coincident visible and thermal observations acquired by THEMIS. We recall here the description of the image processing at visible and infrared wavelengths provided by \citeA{Lange2022a}.  The THEMIS visible camera has five filters with band centers located at 425 (band 1), 540 (band 2), 654 (band 3), 749 (band 4), and 860~nm (band 5) \cite{Christensen2004}. We use colorized images that are either conventional "RGB" composite resulting from band 4 (or, if not available, band 3), band 2, and band 1 in the blue channel \cite{Murray2016, Bennet2018}; or "R2B" images, a colorized product where of band 4 (red) and band 1 (blue) are combined using 0.65~$\times$~band~1~+~0.35~$\times$~band~4 to generate a simulated green band used for the RGB composite \cite{Murray2016, Bennet2018}. 

Surface temperatures are derived from THEMIS band 9 centered at 12.57~\mum~because of its good signal-to-noise ratio and because it is low sensitivity to atmospheric effects \cite{Fergason2006, Pilorget2013}. The precision of the measurement is $\sim$0.5~-~1~K \cite{Christensen2004} with an absolute accuracy of $\sim$2.8~K at 180~K \cite{Fergason2006}, a common temperature for water frost (see section \ref{ssec:frosttemp}). We use the Projected Brightness Temperature (PBT) product provided on the Planetary Data System (PDS) \cite{PDSThemisPBT}. No atmospheric correction is made here. However, as shown in section \ref{ssec:frosttemp}, most of the images are taken between 6 a.m. and 9 a.m. and 4 p.m. and 8 p.m., when the contrast between atmospheric temperature and the surface is small, and thus when the atmospheric correction should be small.

THEMIS visible wavelength images are characterized by a footprint of $\sim$18~km  wide on the ground (18 m resolution), smaller than the $\sim 32$~km wide swath of the infrared data. Therefore, infrared data are cropped to the extent of the overlapping visible wavelength data. Infrared images display a resolution of 100~m/pixel. Visible images are therefore "degraded" by performing a bilinear interpolation of the visible image grid to the infrared image grid. 

\begin{figure}[hbt!]
 \centering
 \includegraphics[scale = .4]{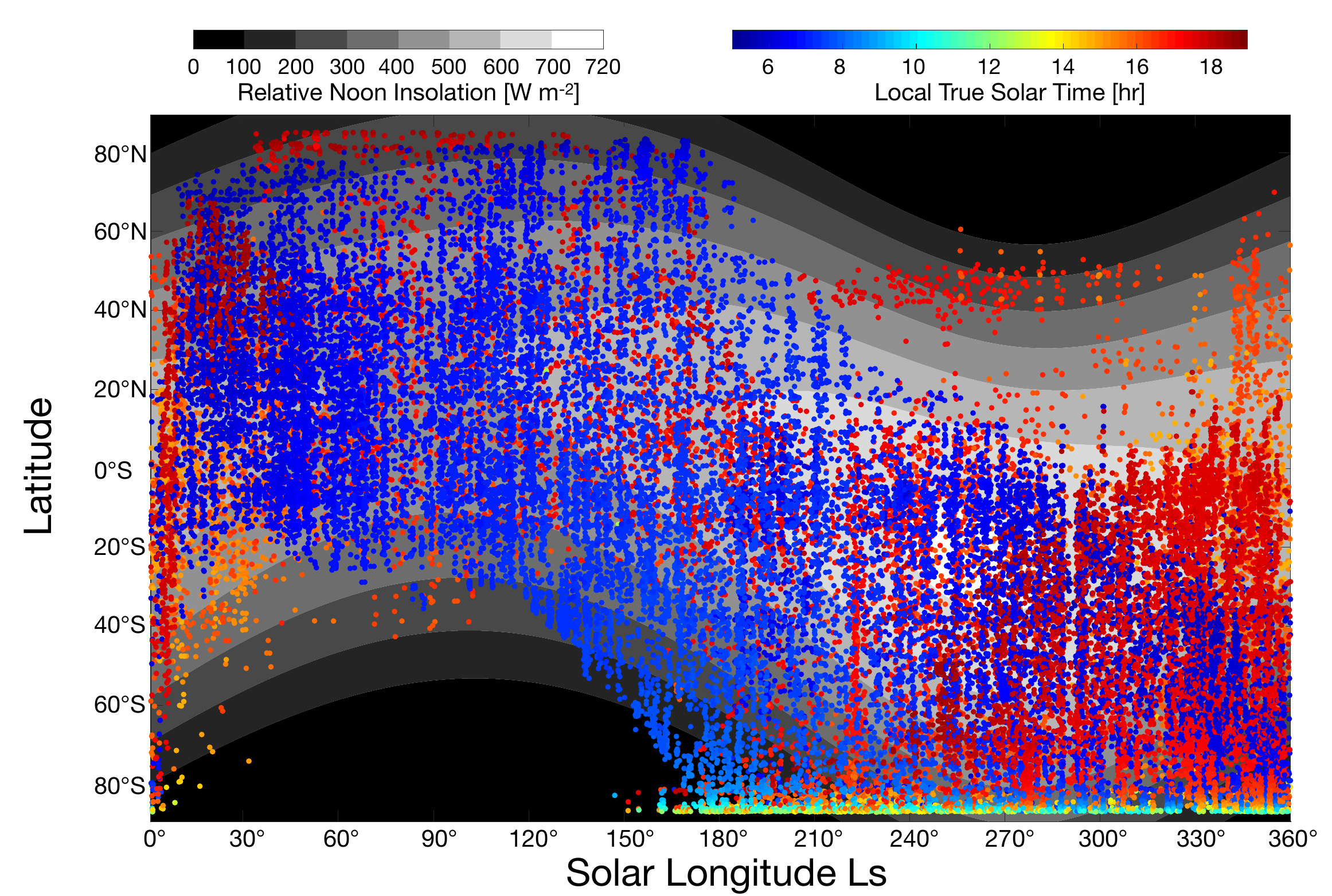}
 \caption{Spatial and seasonal distribution of THEMIS visible/thermal infrared pairs analyzed in this study. The black-white background indicates the seasonality of direct solar illumination (top of the atmosphere) as calculated by a Keplerian orbital model used by KRC \cite{Kieffer2013}. The x-axis represents the solar longitude \Ls~[\textdegree], the Mars-Sun angle, measured from the Northern Hemisphere spring equinox where \Ls~=~0\textdegree.} 
 \label{fig:repartition_alldataset}
\end{figure}

Data analyzed here have been acquired during the daytime, with a local time varying from 5 a.m. to 7 p.m., to have a clear image of the surface at visible wavelengths. We only analyze the visible images acquired concomitant to measurements at infrared wavelengths. Some image pairs are disqualified because of 1) calibration issues in both visible and infrared wavelengths, most likely due to the challenging illumination conditions, and 2) unclear surface exposure due to the presence of clouds/dust. The complete dataset, presented in Figure \ref{fig:repartition_alldataset}, represents 39,496~images.

The dataset analyzed in this study presents some bias caused by operational, local weather, and illumination constraints (e.g., fewer images are taken in the northern latitudes during the second part of the Martian Year [MY] as the sky is dustier \cite{MONTABONE201565}, challenging the observations of the surface). The dataset exhibits a strong asymmetry in the coverage of the Northern and Southern Hemispheres, thus making a comparison between the two hemispheres difficult. Also, few measurements were made during the autumn/winter at mid and high latitudes, preventing the study of condensation of water frost during this period.

\subsection{Water Frost Identification}
\label{ssec:wfrostidentification_method}

The identification of water frost is done in two steps, beginning with looking at visible images. Snow or glacier water ice is associated with a high albedo at visible wavelengths, with a spectral slope from blue to red, in contrast to the low albedo of the Martian bare surface \cite<>[and references therein]{Putzig2005, Singh2016, Singh2018, Murchie2019, Flanner2021, Khuller2021albeo}.  For this reason, we similarly anticipate pure CO$_2$ or H$_2$O ice to appear as blue-white patches on the brown Martian surface. Hence, we first look at all the visible images and manually flag all the pixels that appear blue-white and which could be interpreted as frost. One of the key limits of this approach is that some thin frost layers might not appear blue-white on THEMIS images and would not be detected with our method. Here, we assume that water frost appears blue-white if its thickness is at least 20 micrometers. Such thickness is consistent with the work of \citeA{Svitek1990}, who showed that the white surfaces appearing on Viking images are associated with such frost thicknesses, and with the laboratory work of \citeA{Yoldi2021}, who showed that a thickness of 20~–~30 microns of water frost quintuples the reflectivity of the surface in the blue band and doubles it in the red band (see their Figure 7). A similar detection threshold was also found in \citeA{Spadaccia2023}. We acknowledge that this empirical threshold limit should also depend on the ice crystals' size, the frost's dust content, the emission angle of the observations, etc. \cite{Pommerol2013}. This might explain why our threshold is underestimated: as shown in section \ref{sssec:distributionspatial}, according to the Mars PCM, frosts detected with THEMIS are mostly thicker than 100~\mum~(with a few frosts in the 20 and 100-micron range). 

Although frost/ice detection based on the bright blue-white color contrasting with the bare surface has been widely used in the past \cite<e.g.,>{James1979, Schorghofer2006, CALVIN2015North, BAPST2015, Dundas2019gullies, Lange2022a}, this technique, which is also used in this study, has some important limitations worth mentioning. First, \citeA{Khuller2021albeo} have shown that the albedo of water snow/ice can be drastically reduced if it is contaminated by dust (e.g., less than 1\% dust contamination). Hence, dusty frost could be missed with this approach because of the small contrast between this dirty ice and the bare ground. To mitigate this effect, we have stretched the visible images to enhance the color contrast and to help the detection of small patches of frost in relatively low-illuminated areas (e.g., pole-facing slope during the autumn). Second, as noted by \citeA{Dundas2019gullies} and \citeA{Lange2022a}, relatively blue lithic material can be misunderstood as frost with our approach. While most of the blue-white units we identified are confidently attributed to ice based on their sharp boundaries following topography, preferential slope orientation, or morphology, we leveraged the few uncertain detections by looking at summer images taken at the same locations to see if the blue-white patches were still present. Another possibility was to confirm the icy nature of these pixels by performing an analysis of the spectral properties with all THEMIS bands \cite<as made for instance by>[with the High-Resolution Imaging Science Experiment dataset]{Khuller2021ssice}, but the large number of pixels analyzed here prevent a manual check of all the spectra. Finally, haze/clouds, which can also appear as blue-white on THEMIS visible images \cite<see for instance>{McConnochie2010} sometimes induce false positive detections. However, in this case, these bright features appear to be independent of the surface topography, which enables their removal with confidence. 

To flag pixels as frost, based on their color, we first apply an initial filter to isolate all pixels with a blue-white color. This method eliminates most of the non-frosted pixels. Next, ambiguous pixels (e.g., bright pixels on a slope exposed to sunlight that appear yellow-white) are eliminated manually. Finally, a last check is made between the raw image and the selected pixels to ensure that no frosted areas have been overlooked. However, we acknowledge that this may result in some frosted pixels being omitted.

\begin{figure}[hbt!]
 \centering
 \includegraphics[scale = .25]{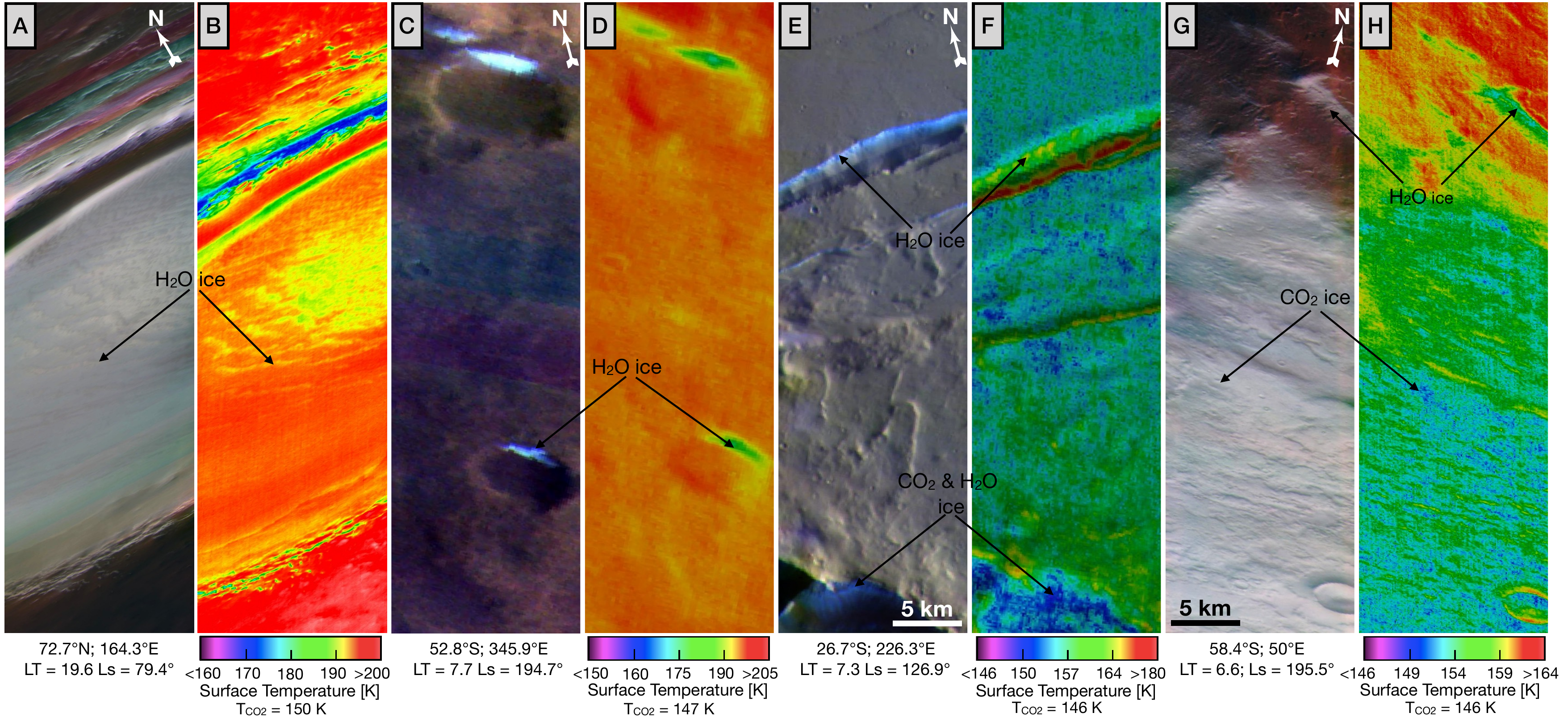}
 \caption{Examples of water ice detections with THEMIS visible wavelength images (a, c, e, g) and corresponding thermal infrared (b, d, f, h) images. Coordinates, solar longitude \Ls, local true solar time (LTST), and temperature of condensation of CO$_2$ are given in the different panels. Blue-white pixels in visible wavelength images with a temperature that is green/yellow/red in the infrared images can be associated with water ice. The white arrows on the upper right of each visible-wavelength image point to the North. Thermal infrared images are underlain with a MOLA background to enhance topography \cite{Zuber1992}. Some terrains appear black in the thermal infrared images because of the background mosaic (not because of an absence of measurement). a-b) Extract from the images V62050008/I62050007, consistent with perennial water ice. Water ice pixels on the complete THEMIS image (and not just the extract presented here) have a mean temperature of 199.33~$\pm$~5~K at \sigO. c-d) Extract from the images V64882003/I64882002, identified as seasonal water ice. Water ice pixels on the complete THEMIS image have a mean temperature of 191.3~$\pm$~3.3~K at \sigO. e-f) Extract from the images V63327007/I63327006, where seasonal water ice (upper part of the image) and diurnal CO$_2$ ice (on the crater rim at the bottom) are present. Water ice pixels on the complete THEMIS image have a mean temperature of 158.1~$\pm$~4~K at \sigO. g-h) Extract from the images V56557003/I56557002, where seasonal water ice is present next to the CO$_2$ ice cap (outside the visible image frame). Water ice pixels on the complete THEMIS image have a mean temperature of 155.9~$\pm$~3~K at \sigO. The contrast on visible images has been manually increased to highlight the frost.}
 \label{fig:frost_identification}
\end{figure}

At this step, 2,343 images (and nearly 9$\times$10$^7$ pixels) are flagged as images showing frost. The second step is to now distinguish between CO$_2$ and H$_2$O frost. As CO$_2$ is the main component of the Martian atmosphere, the formation of CO$_2$ frost is not diffusion-limited. As such, CO$_2$ frost forms when the surface temperature reaches the temperature of condensation of \CO, $T_{\rm{CO}_2}$~[K], given by the Clapeyron law \cite{Jammes1992}:

\begin{equation}
T_{\rm{CO}_2}= \frac{3182.48}{23.3494 - \ln(P)}
 \label{Eq:EqTCO$_2$}
\end{equation}

\noindent with $P$ the local CO$_2$ partial pressure taken as 0.96 of the total surface pressure (expressed in mbar) derived from the local topography and parameterized surface pressure observations \cite{Withers2012}. $T_{\rm{CO}_2}$ ranges from 130~K at the top of Olympus Mons to more than 153~K in Hellas basin \cite{Piqueux2016}. If CO$_2$ ice is present on the surface, the temperature must be at $T_{\rm{CO}_2}$. On the contrary, the formation of H$_2$O frost is limited by diffusion and is controlled by the partial pressure of H$_2$O at the surface and the near-surface atmosphere. Water frost can thus exhibit a surface temperature much higher than $T_{\rm{CO}_2}$. Therefore, for each pixel flagged as frost, we compare its surface temperature given by the THEMIS infrared measurement with $T_{\rm{CO}_2}$ (Figure \ref{fig:frost_identification}b). To account for the instrument noise and possible atmospheric contributions, we assign a 5~K tolerance on this criterion as in \citeA{Lange2022a}. In other words, a pixel is considered as water frost if its temperature $T_{\rm{ice}}$ is higher than $T_{\rm{CO}_2}$~+~5~K. In comparison, \citeA{Pilorget2013} used a 6~K margin, and \citeA{Khuller2021albeo} a 7~K margin. Using these tolerances yields a reduction of the number of water ice detections by 1.6\% and 3.8\% respectively. With this approach, we exclude water frost at very low temperatures that might be mixed with CO$_2$ frost and thus consider only “pure” water frost (although dust might be incorporated). For this reason, frost/ice will now only refer to water frost/ice in the rest of this manuscript. 

Occasionally, some temperature measurements are associated with large uncertainties \cite<e.g., all measurements above 170~K on the seasonal CO$_2$ ice cap,>{Wagstaff2008}. This overestimation is well-characterized and linked to the image calibration protocol that uses a reference image unsuitable at high latitude during winter / early spring \cite<see a complete description of this issue in>{Wagstaff2008},  and has been mitigated with the new calibration of PDS products since their work. Such images, where the CO$_2$ ice cap is associated with unrealistic high temperatures (here taken as 160~K) are manually removed from our dataset.

At the end of the processing, 2,006 images (and 5.3$\times$10$^7$ pixels) are found to be associated with water frost (e.g., Figure \ref{fig:frost_identification}). The reader is referred to \citeA{Khuller2021frost} and \citeA{Lange2022a} for a mapping of CO$_2$ frost with the THEMIS dataset. As a preliminary validation, we have checked that perennial water ice deposits (North Cap, Korolev crater, Lyot crater, etc.) are well identified by this method (e.g., Figures \ref{fig:frost_identification}a, b). The good agreement between our detection method and the other datasets, as well as the PCM, allows us to be confident in the validity of our detection method (section \ref{sec:results}).

\subsection{The Mars Planetary Climate Model}

Observational data are compared to the simulations from the Mars Planetary Climate Model, formally known as the LMD Mars Global Climate Model \cite{Forget1999}. In this study, we use the version that models slope microclimates and simulates the condensation of CO$_2$ and H$_2$O on slopes \cite{Lange2023Model}. Surface properties (albedo, emissivity, thermal inertia) are set to the observations from TES \cite<see Table 2 of >{Lange2023Model}. Seasonal dust opacity profiles are set to an average of the available observations of dust from MY 24, 25, 26, 28, 29, 30, and 31 outside the global dust storm period \cite{MONTABONE201565}. The representation of the water cycle, detailed in  \citeA{Navarro2014} and \citeA{Naar2021}, has been validated through comparison with TES data. A complete description of the surface energy budget, accounting for visible and infrared radiation, soil conduction, and sensible and latent heat fluxes can be found in \citeA{Lange2023Model}.

\subsubsection{Modeling the formation and sublimation of water frost}
\label{ssec:modelh2ofrost}
Three kinds of models are used to simulate the evolution of water ice \cite<see a complete review in>{Khuller2024}. One of the most common approaches \cite<e.g.,>{Williams2008,Dundas2010,Kite2013,Bapst2019,Bramson2019} is to sum the sublimation of water ice induced by the buoyancy of near-surface air, especially given the contrast of mass between the CO$_2$ gas (molar mass of 44~g~m$^{-3}$) and H$_2$O (molar mass of 18~g~m$^{-3}$) \cite{Ingersoll1970,Schorghofer2020, Khuller2024}, known as "free convection" (referred to below as "water buoyancy"), and the one driven by wind advection, known as "forced convection". Yet, as shown by \citeA{Khuller2024}, these kinds of models have several limitations, including: 

\begin{itemize}
    \item wind-shear and buoyancy terms should not be summed, since they represent two distinct atmospheric regimes. As such, the sublimation rate may be overestimated when summing these two terms;
    \item ignoring the physical processes in the near-surface and interfacial layers. In the interfacial layer, viscous effects reduce the sublimation rate. While it has a significant effect on Earth \cite{Fitzpatrick2019}, almost none of the models used on Mars today account for this effect. In the surface layer, gustiness induced by large-scale transient convective structures under convective conditions adds shear stress which can promote the sublimation of water ice. Thus, even in the absence of a geostrophic/local wind, the "free convection" regime cannot be achieved, as convective cells will generate a gustiness wind close to the surface.
\end{itemize}

Hence, as shown by \citeA{Khuller2024}, models which sum the wind-shear and buoyancy terms, and which neglect the physical processes in the near-surface and interfacial layers may inaccurately estimate sublimation rates. Here, we simulate the evolution of water ice using a common approach in near-surface atmosphere modeling known as the "bulk method" \cite{Flasar1976, Savijrvi1995, Montmessin2004, Colaitis2013,Navarro2014, Haberle2019}, which relies on the Monin-Obukhov theory widely validated on Earth \cite{Foken2006},  and which can be transposed on Mars \cite{Martinez2009}.

 The evolution of the mass of H2O frost $m_w$~(kg~m$^{-2}$) is computed with:
\begin{equation}
     \frac{\partial m_{w}}{\partial t}= \rho C_q U (q_w - q_{sat}(T_{\rm{surf}}))
\end{equation}
\noindent where  $\rho$~(kg~m$^{-3})$ is the air density, $U$ (m~s$^{-1}$) is the wind velocity obtained by combining the large-scale (synoptic) wind near the surface with a wind gustiness induced by buoyancy \cite{Colaitis2013}, $q_{w}$~(kg/kg) is the mass mixing ratio of water vapor in the first layer of the model ($z_1 \approx 4$~m), $q_{sat}$~(kg/kg) is the saturation mass mixing ratio computed from the surface temperature (see Eq. (1) of \cite{Pal2019}). $C_q$ (unitless) is a moisture transfer coefficient  given by \cite{Colaitis2013}:
\begin{equation}
    C_q = f_q(Ri) \left( \frac{\kappa^2}{\ln{\frac{z_1}{z_0}}  \ln{\frac{z_1}{z_{0q}}}}\right)
\end{equation}

\noindent where $f_q(Ri)$~(unitless) is a function of the Richardson number $Ri$~(unitless),  $\kappa$~(unitless) is the von Kármán constant set to 0.4; $z_0$~(m) is the aerodynamic roughness coefficient extracted from \citeA{Hebrard2012},  $z_{0q}$ (m) the moisture roughness length. $f_q(Ri)$ amplifies the flux for an unstable atmosphere, while it reduces it for a stable atmosphere. The Richardson number $Ri$ is the ratio of the buoyancy term to the flow shear term, and depicts the stability of the atmosphere: 
\begin{equation}
    Ri = \frac{g}{\theta_{s}} \frac{(\theta - \theta_s)z}{U^2}
    \label{eq:turb_rib}
\end{equation}

\noindent where $g$~(m~s$^{-2}$) is the gravity field, $\theta_{s}$~(K) is the surface temperature, $\theta$~(K) is the potential temperature at altitude $z$~(m). $Ri = 0$ for a stable atmosphere, $Ri < 0$ for an unstable atmosphere, $Ri > 0$ for a stable atmosphere. The stability functions $f_q(Ri)$ used are the ones from \citeA{England1995}. The stability functions used for the moisture are the same as for the heat as it is a consistent assumption validated with Earth experiments \cite<e.g.,>[and references therein]{Jimnez2012}. These functions have been well-validated on Earth, in Antarctica \cite{Vignon2016}, and have significantly improved the predictions of the near-surface environment on Mars, especially during the daytime,  when the near-surface atmosphere is unstable \cite{Colaitis2013}.

$z_{0q}$, which represents the viscous effects in the interfacial layer,  is parameterized as the thermal roughness length  given by  \citeA{Brutsaert1982}:

    \begin{equation}
    z_{0q} = z_0\exp(-7.3 \kappa Re^{1/4} Pr^{1/2}+5\kappa)
        \label{eq:z0tc}
\end{equation}

\noindent where $Re$~(unitless) is the Reynolds number derived from the wind, density, and viscosity given by the model in the near-surface layer, $Pr$~(unitless)  the Prandtl number derived from kinematic viscosity and thermal diffusivity in the model. The application of this parameterization to the Martian surface layer has been discussed by \citeA{Martinez2009}.  In our simulations,   $z_{0q}$ is nearly 1/10 of $z_0$. Given the low Reynolds number on Mars (global -over the planet- mean value of 1.7 over the year in our simulations),  the approximation that the moisture roughness length is equal to the thermal roughness length is relevant \cite{Larsen2002}.

One of the key assumptions in our model is that the stability of the atmosphere only depends on the thermal contrast between the atmosphere and the surface, and the instability induced by the difference of molar weight between H$_2$O and CO$_2$ \cite{Ingersoll1970, Schorghofer2020, Khuller2024} is not considered in the stability functions and the computation of the gustiness. We show that this approximation is reasonable for our study below.

\subsubsection{Effect of water buoyancy on the atmospheric stability and the sublimation rate}

To account for the effect of water buoyancy in the overall buoyancy in the near-surface layer, the definition of the Richardson number $Ri$ (noted $Ri_{dry}$  in this case) can be modified with: 
\begin{equation}
    Ri_{wet} = \frac{g}{\theta_{vs}} \frac{(\theta_v - \theta_s)z}{U^2}
    \label{eq:turb_rib}
\end{equation}

\noindent $\theta_v$~(K) is the virtual potential temperature defined by:
\begin{equation}
\theta_v = \frac{\theta}{ 1-\frac{P_v}{P}(1-\epsilon)}
\end{equation}
\noindent where $P_v$~(Pa) is the partial pressure of water vapor, $P$~(Pa) the total pressure, $\epsilon$ the ratio of the molar mass of water vapor v.s. dry air (i.e., CO$_2$ gas). This quantity represents the temperature that dry air with the same density and pressure as moist air would have. In case ice is present at the surface or at an altitude $z$, $P_v$ is given by the saturation pressure over ice \cite{Murphy2005}.

\begin{figure}[hbt!]
 \centering
 \includegraphics[width = \textwidth]{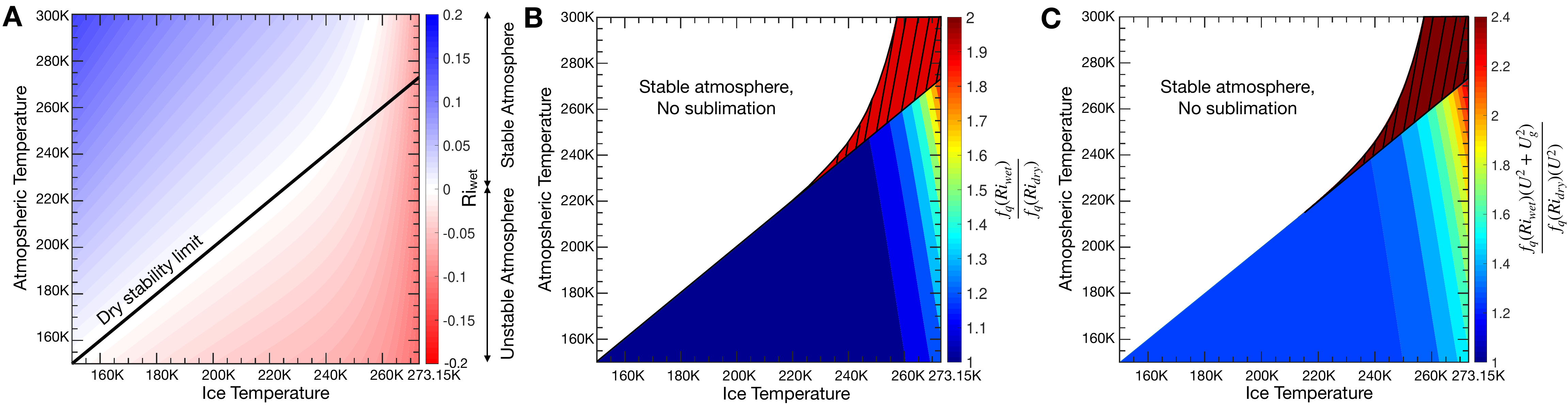}
 \caption{a) Wet Richardson number as a function of the atmospheric temperature (at 4~m, first atmospheric layer of the Mars PCM) and surface temperature. A negative Richardson number (presented in red in the Figure) indicates an unstable atmosphere. The dry stability limit ($Ri_{dry}$~=~0) is indicated by the dark curve. b) Ratio of the stability function computed with the wet Richardson number and the dry Richardson number. The dashed areas correspond to a stable atmosphere according to the dry Richardson number, where the stability function $f_q(Ri_{dry})$~=~0. In this area, $f_q(Ri_{wet})$ ranges between 1 and 1.2. c)  Ratio of the stability function computed with the wet Richardson and the near-surface wind speed accounting for gustiness induced by water buoyancy \cite{Colaitis2013,Khuller2024} with the stability function computed with the dry Richardson and the near-surface wind speed. Dashed lines is not considered for the same reason as in b). }
 \label{fig:buyoancy}
\end{figure}

We present in Figure \ref{fig:buyoancy}a the wet Richardson number as a function of the surface and atmospheric temperatures, with parameter $U$~=~10~m~s$^{-1}$ \cite<a typical value during daytime,>{Martnez2017}, $P$~=~610~Pa, the global averaged Martian Pressure, and no water vapor in the atmosphere to maximize the effect of water buoyancy. No significant changes happen when changing the atmospheric humidity from 0 to the current humidity. The stability limit is given by $Ri_{wet}$~=~0 (white part in Figure \ref{fig:buyoancy}a). The dark curve represents the "dry" stability limit (i.e., without the effect of water buoyancy). The wet stability limit at low temperatures agrees well with the dry limit at low temperatures and starts to deviate at $\sim$240~K. This limit changes slightly by a few kelvins when varying the total pressure $P$). At high temperatures ($\ge$260~K), the effect of water buoyancy induces a strong instability and surpasses the strong, stable stratification between the warm atmosphere and colder surface. This is mainly due to the exponential dependence of saturation pressure over water ice  \cite<Eq. \ref{eq:murphypsv},>{Murphy2005}, which at warm temperatures, can reach the average surface pressure on Mars.  The difference in wind gustiness when accounting for this effect is $\sim$5~m~s$^{-1}$ at 273~K \cite{Khuller2024}.

To quantify the effect on the sublimation rate, we first compare the product $f_q(Ri_{wet})$ and $f_q(Ri_{dry})$, without accepting for gustiness, in Figure \ref{fig:buyoancy}b. The stability function $f_q$ is barely affected by the effect of water buoyancy, except at high temperatures close to the melting point of water (273.15~K). The buoyancy is mostly dictated by the temperature contrast between the surface and near-surface atmosphere.  Yet, when accounting for a gustiness wind U$_g$~=~5~m~s$^{-1}$ induced by the water buoyancy \cite{Khuller2024}, the sublimation rate can increase by 25\% (Figure \ref{fig:buyoancy}c) at low temperatures (which is a bit unrealistic since the atmospheric instability is very weak at these low temperatures, Figure \ref{fig:buyoancy}a, and the gustiness induced by water buoyancy should be thus much lower), and ~100\%  close to the melting point. Hence, our approximation to neglect the effect of water buoyancy in our simulation is correct a low temperatures ($\le$240~K). As shown in section \ref{ssec:frosttemp}, this is the range of water ice temperatures modeled and measured on Mars. Hence, our results should not be significantly impacted by our assumption to neglect water buoyancy.

 \subsubsection{Model validation}

\begin{figure}[hbt!]
 \centering
 \includegraphics[scale = .35]{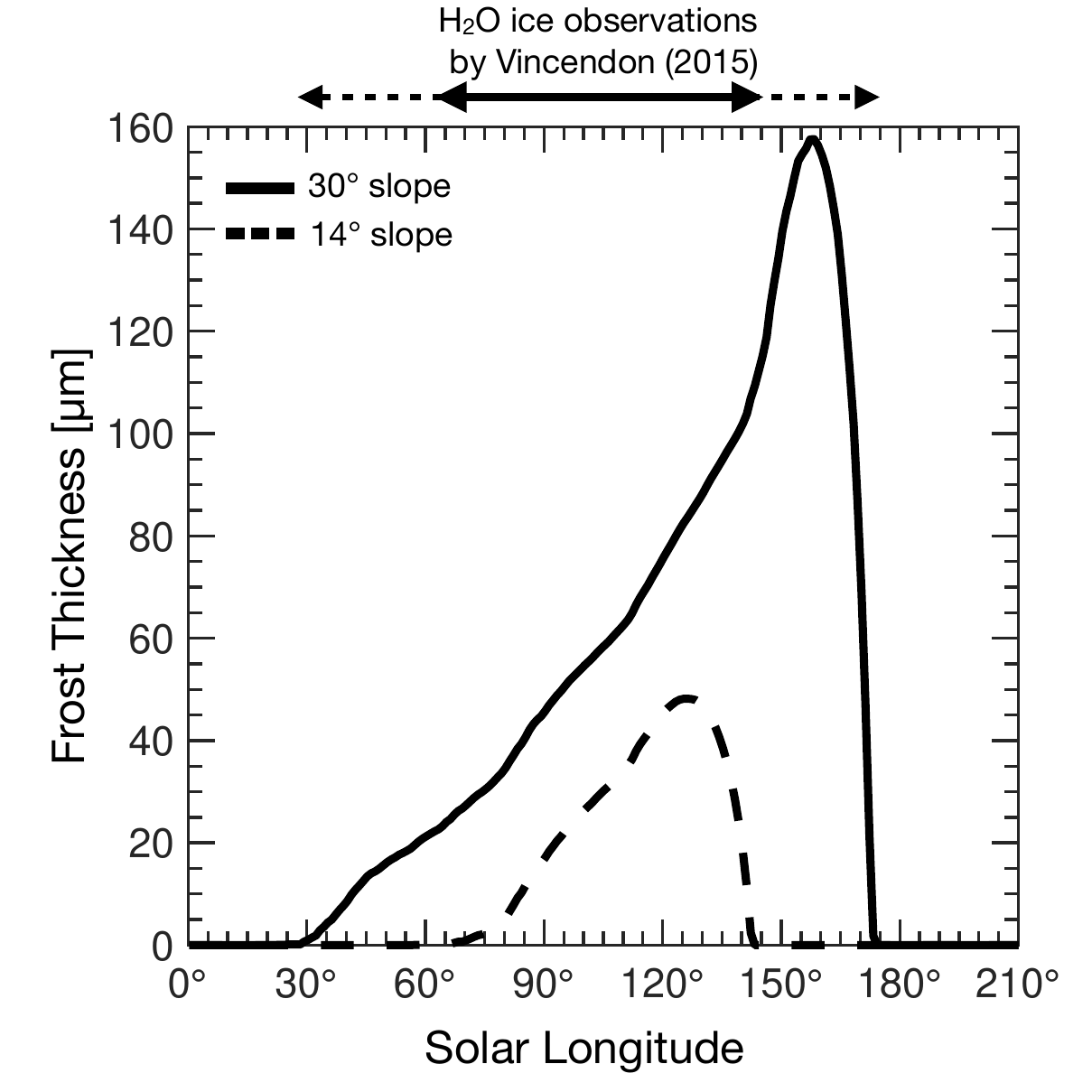}
 \caption{Water frost thicknesses predicted by the Mars PCM at Promethei Terra (32.3\textdegree S, 118.6\textdegree E) compared to the observations (plain arrow) of water frost by \citeA{Vincendon2015} (uncertainty in the detections presented with the dashed arrow).}
 \label{fig:FrostThickPromete}
\end{figure}

As stated in section \ref{ssec:modelh2ofrost}, the methods used in this model have been widely used and validated on Earth. On Mars, this model has been successfully applied to simulate the sublimation of the massive perennial water ice deposits at the North Pole and the release of water vapor in the atmosphere \cite{Navarro2014}. The model also agrees well with the observations of water ice exposure close to the CO$_2$ seasonal cap \cite{Langevin2007}, and the presence of water ice frost on pole-facing slopes at mid-to-low latitudes \cite{Vincendon2010water,Vincendon2015, Lange2023Model}. The sub-grid slope parameterization has been validated with comparisons between the model outputs and surface temperatures measured on slopes by several instruments \cite{Lange2023Model}. We present in Figure \ref{fig:FrostThickPromete} the H$_2$O frost thickness at Promethei Terra (32.3\textdegree S, 118.6\textdegree E) and compare them with observations of frost on the crater rim with high temporal coverage by Vincendon (2015). The good agreement between the model outputs and the observations \cite<which can be generalized to other locations presented in>[not shown here]{Vincendon2015} validate our method, the modeled timing of formation/sublimation of the frost, as the modeled frost thickness.

\subsubsection{Model limitations}
There are two main limitations in the model used. First, the effect of water buoyancy should be considered. This requires modifying the parameterization of the Planetary Boundary Layer in the Mars Planetary Climate to account for convection driven by H$_2$O buoyancy \cite{Ingersoll1970,Schorghofer2020,Khuller2024}, CO$_2$ buoyancy at high latitudes in the polar night \cite{Hess1979}, and include possible phases changes / latent heat released. This is left as a future work. Furthermore, on slopes, strong slope winds can have a significant impact on surface temperatures \cite{Spiga2018,Smith2018}, and thus should fasten the sublimation of ice. A parameterization of the slope winds in the Mars PCM is under implementation.

\section{Results \label{sec:results}}
\subsection{Distribution of Frost}

\subsubsection{Spatial Distribution}
\label{sssec:distributionspatial}

\begin{figure}[hbt!]
 \centering
 \includegraphics[scale = .35]{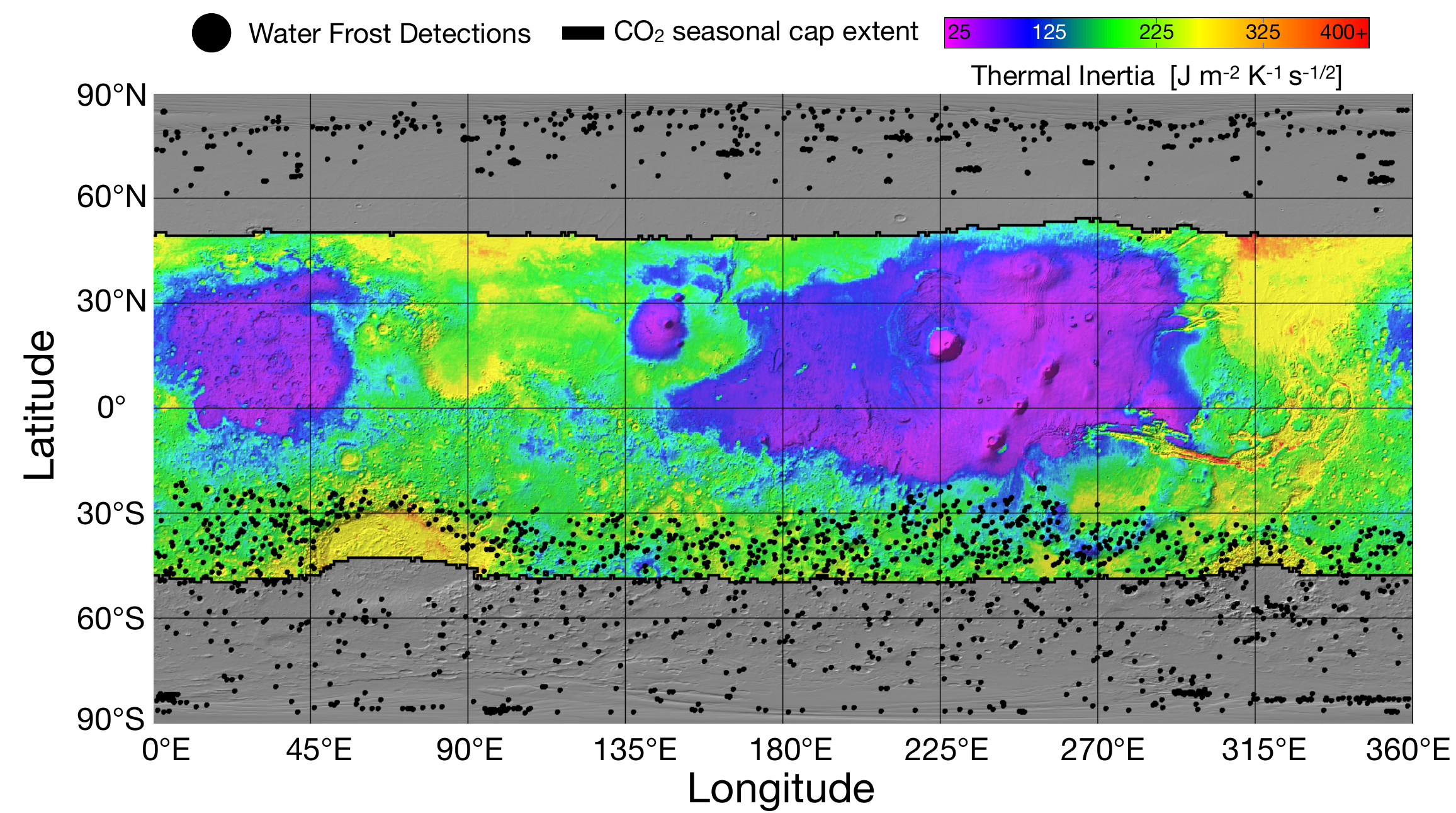}
 \caption{Distribution of water frost detected with THEMIS (black dots). The colorized background is a thermal inertia map from \citeA{Piqueux2023timap} overlain by a MOLA-shaded relief \cite{Zuber1992}, only shown outside the maximum extent of the continuous seasonal caps \cite{Piqueux2015}}
 \label{fig:distribution_spartial_frost}
\end{figure}

The spatial distribution of frost is presented in Figure \ref{fig:distribution_spartial_frost}. Water ice is detected down to 21.4\textdegree S, 48.4\textdegree N. 91\% of the low-latitude detections (in the $\pm$45\textdegree N band) occur on pole-facing slopes, where water vapor preferentially condenses \cite{Vincendon2010water, Lange2023Model}. At higher latitudes, 64\% of the detections are made on pole-facing slopes. In comparison, \citeA{Carrozzo2009} and \citeA{Vincendon2010water}  have detected water frost on pole-facing slopes down to 15\textdegree S and 13\textdegree S–32\textdegree N respectively using near-infrared data from OMEGA (Observatoire pour la Minéralogie, l'Eau, les Glaces et l'Activité) and CRISM (Compact Reconnaissance Imaging Spectrometer for Mars) onboard the Mars Reconnaissance Orbiter.

\begin{figure}[hbt!]
 \centering
 \includegraphics[scale = .35]{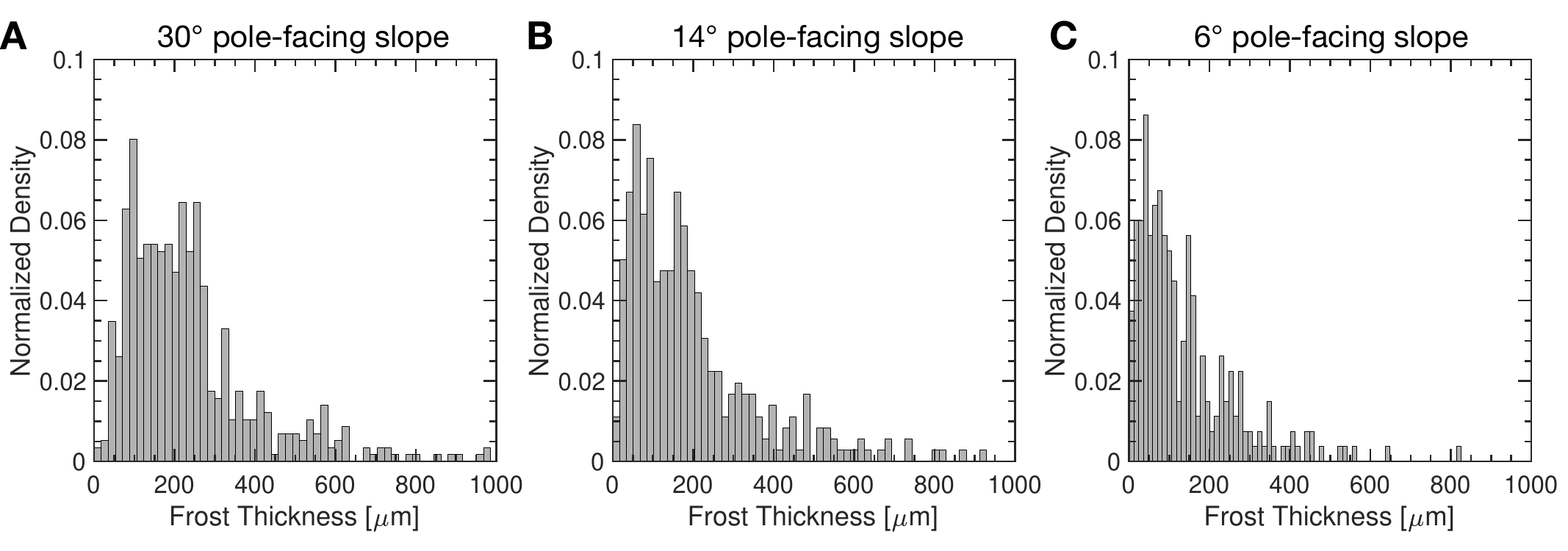}

 \caption{a) Distribution of water frost thickness detected with THEMIS between 20\textdegree S and 50\textdegree S as modeled by the Mars PCM. Since most of the detections are made on the top of crater rims, we have derived the frost thicknesses assuming a pole-facing slope of 30\textdegree. Cases with modeled frost thickness larger than 1000~\mum~not shown. For completeness,  the distribution of water frost thicknesses for a 14\textdegree~slope (b) and 6\textdegree~slope (c) is presented. The case for a flat surface is not presented as almost all of the frosts detected in this latitudinal range are on pole-facing slopes. }
 \label{fig:frost_thick}
\end{figure}

The difference in the latitudinal extent of the frost detected with THEMIS and OMEGA/CRISM can be explained by the intrinsic properties of each dataset and instrument. We have estimated the thickness of the frost detected with THEMIS on crater slopes between 50\textdegree S and 20\textdegree S using the PCM (Figure \ref{fig:frost_thick}). THEMIS detects water ice with a median thickness of 180~\mum~(minimum thickness estimated to be near 18~\mum). Hence, our approach mainly detects thick frost layers ($\sim$100~\mum~thick), and our detection threshold estimated in section \ref{ssec:wfrostidentification_method} may have been underestimated. In comparison, OMEGA and CRISM have an ice detection threshold of 2-5~\mum~\cite{Vincendon2010water, Vincendon2015}. At latitudes 10\textdegree S–20\textdegree S, water ice should be thin \cite<a few tens of microns, see Figure 12 of >{Lange2023Model}. Hence, THEMIS does not seem to be able to detect such thin ice at these low latitudes compared with OMEGA/CRISM. 

Between 20\textdegree~and 50\textdegree~latitude, the geographic distribution of water frost is consistent with OMEGA/CRISM observations \cite{Vincendon2010water}. In the South, between 20\textdegree S and 30\textdegree S, two frost-free areas with no frost are found between 100\textdegree E and 210\textdegree E and 250\textdegree E and 360\textdegree E, one of which not observed by \citeA{Vincendon2010water}. Most of the THEMIS observations obtained in this longitudinal band were acquired before \Ls~=~100\textdegree, where frost is expected to be very thin \cite<$\leq$~10~\mum~thick according to models,>{Vincendon2010water, Lange2023Model}  and illumination low, especially on pole facing slopes (Figure \ref{fig:repartition_alldataset}). Hence, frost may be too thin to yield a detectable signature at visible wavelengths, preventing detection with our proposed method. For the 250\textdegree E–360\textdegree E band, two processes may explain the absence of frost. First, the 250\textdegree E–360\textdegree E area is characterized by smooth lava terrains with few slopes \cite{Vincendon2010}, while water frost is only stable on steep pole-facing slopes between 20\textdegree S and 30\textdegree S \cite{Vincendon2010water,Lange2023Model}. Second, for the 300\textdegree E–360\textdegree E area, the absence of frost is a consequence of a drier atmosphere induced by a western jet on the eastern side of Tharsis \cite{Joshi1994,Joshi1995,Vincendon2010water,Lange2023Model}. Most of the deposits found between 20\textdegree S–30\textdegree S are located in the West of Hellas basin, where a southward wet flux of air converges \cite<Figure 7 of>{Vincendon2010water}, promoting the formation of water frost at this location. We also note that no water frost detections are made within the Hellas basin. This is mostly due to a stationary wave induced by the strong topographic depression of Hellas, filling the basin with dry air and thus preventing the formation of water frost \cite{Vincendon2010water}. Challenging conditions (for instance clouds \cite{Langevin2007, Kahre2020}, suspended dust \cite{Martin1993}, etc.) which prevent a clear view of the surface might also explain the absence of frost detections at this location with our method. For instance, \citeA{Langevin2007} have detected some water frost deposits in the South of the basin during spring with OMEGA/CRISM as they do not need a clear exposure of the surface to identify frost. In the North, only one detection is made below 50\textdegree N, but this is a consequence of the bias dataset used in this study and the sparse coverage of the Northern Hemisphere at mid-latitude during northern winter/spring (Figure \ref{fig:repartition_alldataset}). At high latitudes, in both hemispheres, no significant frost-free areas can be found.

\subsubsection{Temporal Distribution}
\label{ssec:Temporaldistribution}
\begin{figure}[hbt!]
 \centering
 \includegraphics[scale = .35]{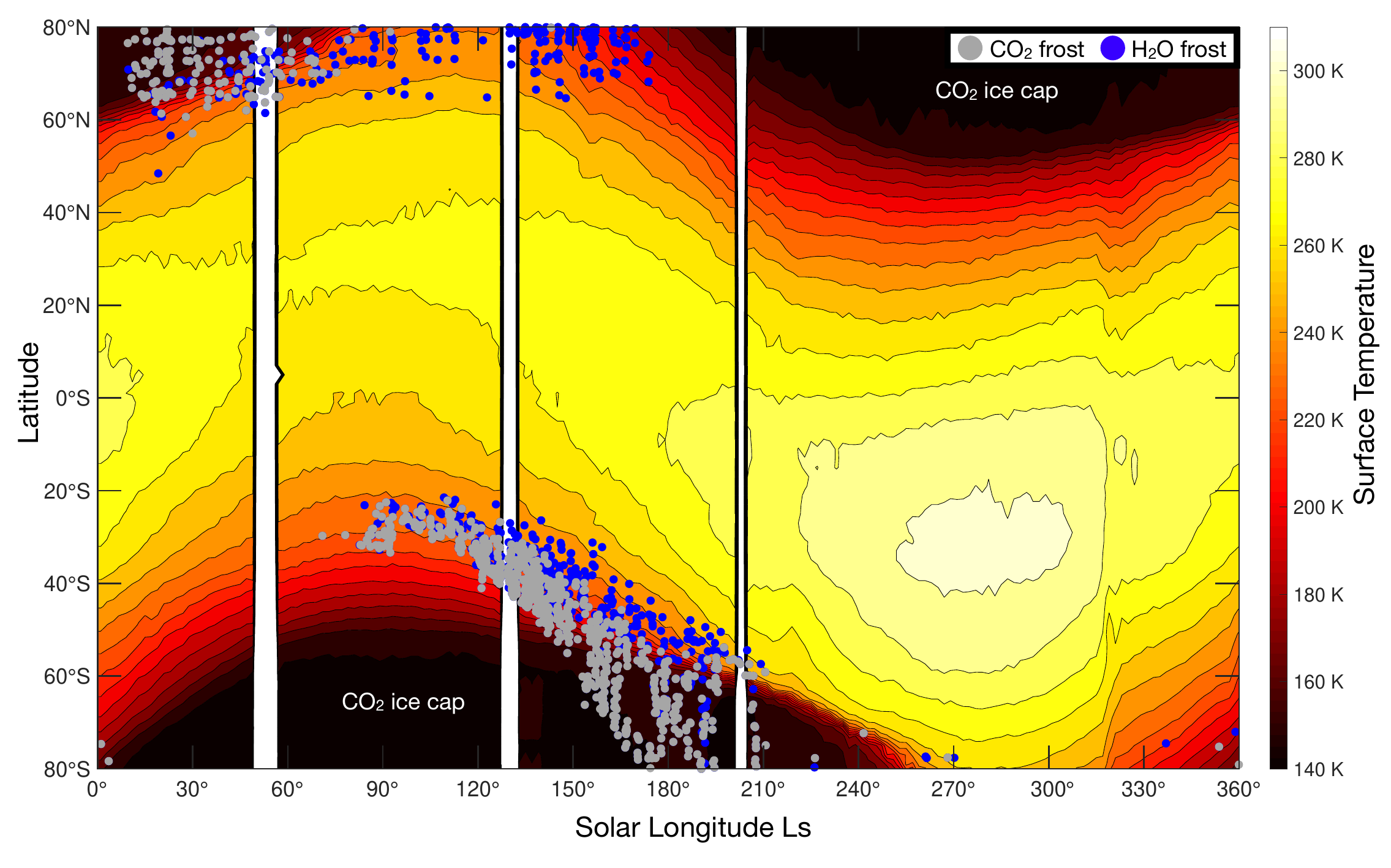}
 \caption{Distribution of water and \CO~frost. The colorized background is the zonal average surface temperature measured by TES at 2~p.m. during MY 26 \cite{Smith2004}. Perennial polar caps at latitudes higher than $\pm$80\textdegree~are not presented here.  }
 \label{fig:distribution_latls_frost_all}
\end{figure}

The temporal distribution of water frost is presented in Figure \ref{fig:distribution_latls_frost_all}. For completeness, we also present CO$_2$ ice frost detected with THEMIS data, i.e., pixels with a temperature lower than $T_{\rm{CO}_2}$~+~5~K. In the South, equatorward of 50\textdegree S, most of the water ice detections are made after the sublimation of the CO$_2$ seasonal ice on pole-facing slopes. Water frost can survive 10–20\textdegree~of \Ls~after the disappearance of CO$_2$ ice. The same comparison is difficult to make in the North, due to the low number of detections. Summer water ice detections are made where perennial water ice is observed (water ice cap at the North Pole, Korolev crater, etc.). 

\citeA{BAPST2015} have exhibited a strong hemispheric asymmetry in the presence of water ice frost during autumn. They showed that widespread water ice deposits can be observed in the North during autumn (before \Ls~$\le$~270\textdegree), but none were detected in the South during southern autumn (before \Ls~$\le$~90\textdegree). We report here 27 occurrences (i.e., 1.3\% of the dataset) of water frost detected during the autumn between 32\textdegree S and 23\textdegree S, mostly in the West of Hellas basin and at longitudes 200\textdegree E–250\textdegree E. All of these detections are made on pole-facing slopes. The same detections have been made by \citeA{Vincendon2010water}. Yet these observations should not contradict \citeA{BAPST2015}'s conclusions since 1) they are made on local sites with favorable thermal conditions (steep pole-facing slopes) and 2) they take place at low latitudes, where \citeA{BAPST2015} were unable to observe ice/frost due to the resolution of TES, far from the CO$_2$ seasonal ice cap where they noted this asymmetry.

\begin{figure}[hbt!]
 \centering
 \includegraphics[scale = .4]{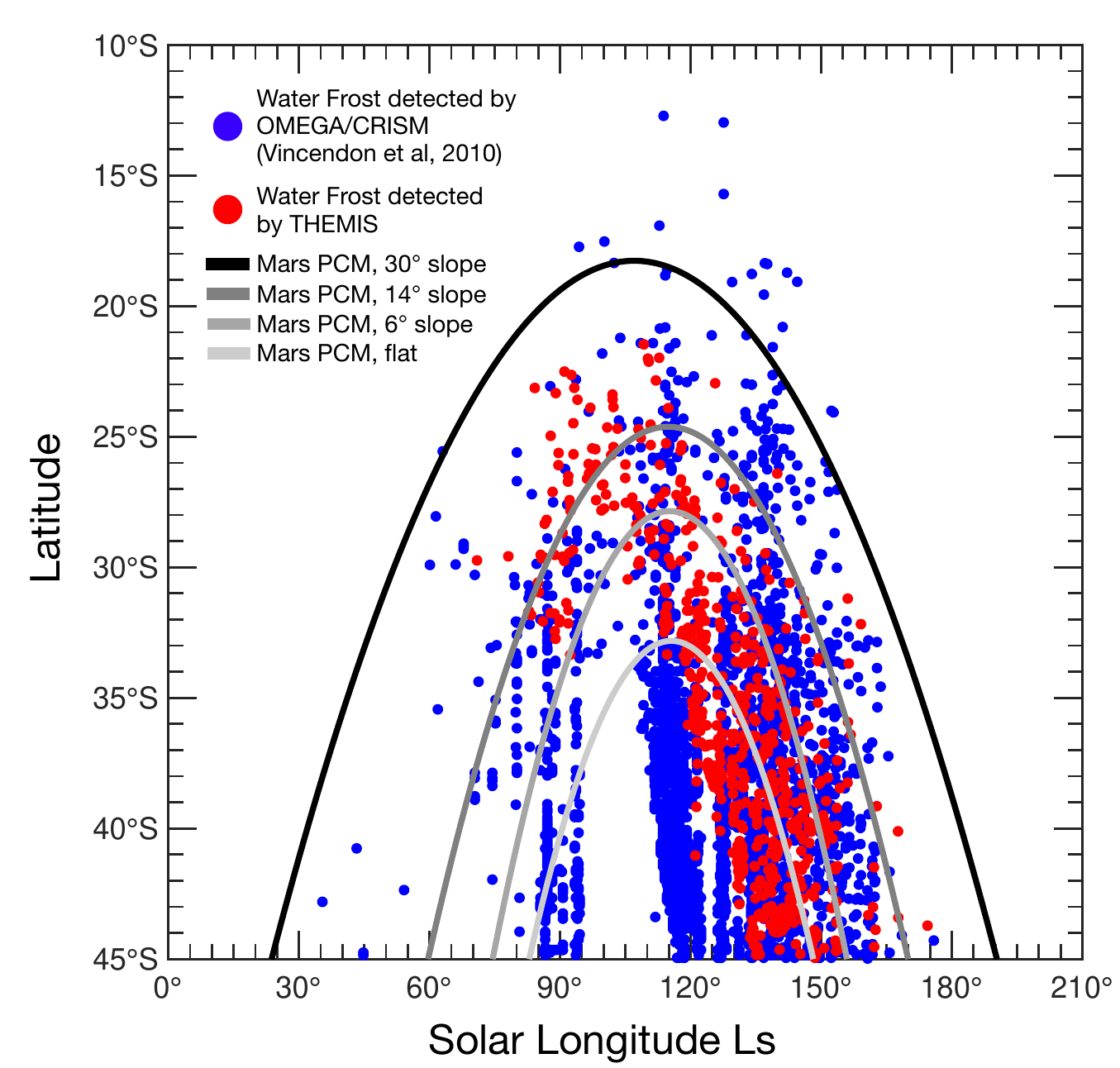}
 \caption{Latitudinal distribution of \HHO~frost versus solar longitude \Ls. Blue points correspond to the observations of frost with CRISM/OMEGA \cite{Vincendon2010water, Vincendon2015}. Red points correspond to the observations of frost with THEMIS. The dark curve is the prediction of frost stability on a 30\textdegree~pole-facing slope retrieved from \citeA{Lange2023Model}, as most of the ice deposits are found on the crater rim where the slopes are the steepest. Other predictions for the slopes modeled in \citeA{Lange2023Model} are given in greys. Following \citeA{Lange2023Model}, \HHO~ice is predicted at a given latitude by the PCM if the \HHO~frost thickness exceeds 5~\mum.  The PCM outputs are retrieved at 2 p.m. to be consistent with OMEGA/CRISM local time of acquisitions. Hence, the dark curve represents the stability of seasonal \HHO~frost on pole-facing slopes.}
 \label{fig:ls_omega_pcm}
\end{figure}

A focus on the water frost evolution at mid-latitudes in the South is presented in Figure \ref{fig:ls_omega_pcm}. Again, this analysis is not performed in the North due to the bias in the distribution of the dataset. THEMIS and OMEGA/CRISM datasets have a similar temporal evolution during southern spring (Figure \ref{fig:ls_omega_pcm}). Few detections have been made with THEMIS during the condensation of H$_2$O (\Ls~$\leq$~120\textdegree) mostly because of the sparse measurements made at this period (Figure  \ref{fig:repartition_alldataset}). Also, at that time, pole-facing slopes are shadowed, preventing a clear exposure of the surface and thus detection of water frost. Finally, ice deposits are expected to be thin during this period (nearly 1-2~\mum). Therefore, this low thickness might reduce the albedo contrast between the frost and the defrosted surface, challenging the detection of frost based on our method. Vincendon et al. (2010) also noted that OMEGA/CRISM could only detect frost with a thickness larger than a few microns, possibly explaining the absence of frost detections at \Ls~$\leq$~80\textdegree. The condensation and sublimation timing of H$_2$O frost deposits detected with THEMIS is consistent with the PCM prediction \cite{Lange2023Model}, although the PCM underestimates the sublimation of H$_2$O by 10\textdegree~of \Ls. It is not clear if the earlier condensation of H$_2$O frost in the PCM is due to a systematic error in the model or a bias due to the sparse number of detections made at that time. 

\subsubsection{Retreat of the Water Frost in the Southern Hemisphere}

\begin{figure}[hbt!]
 \centering
 \includegraphics[scale = .3]{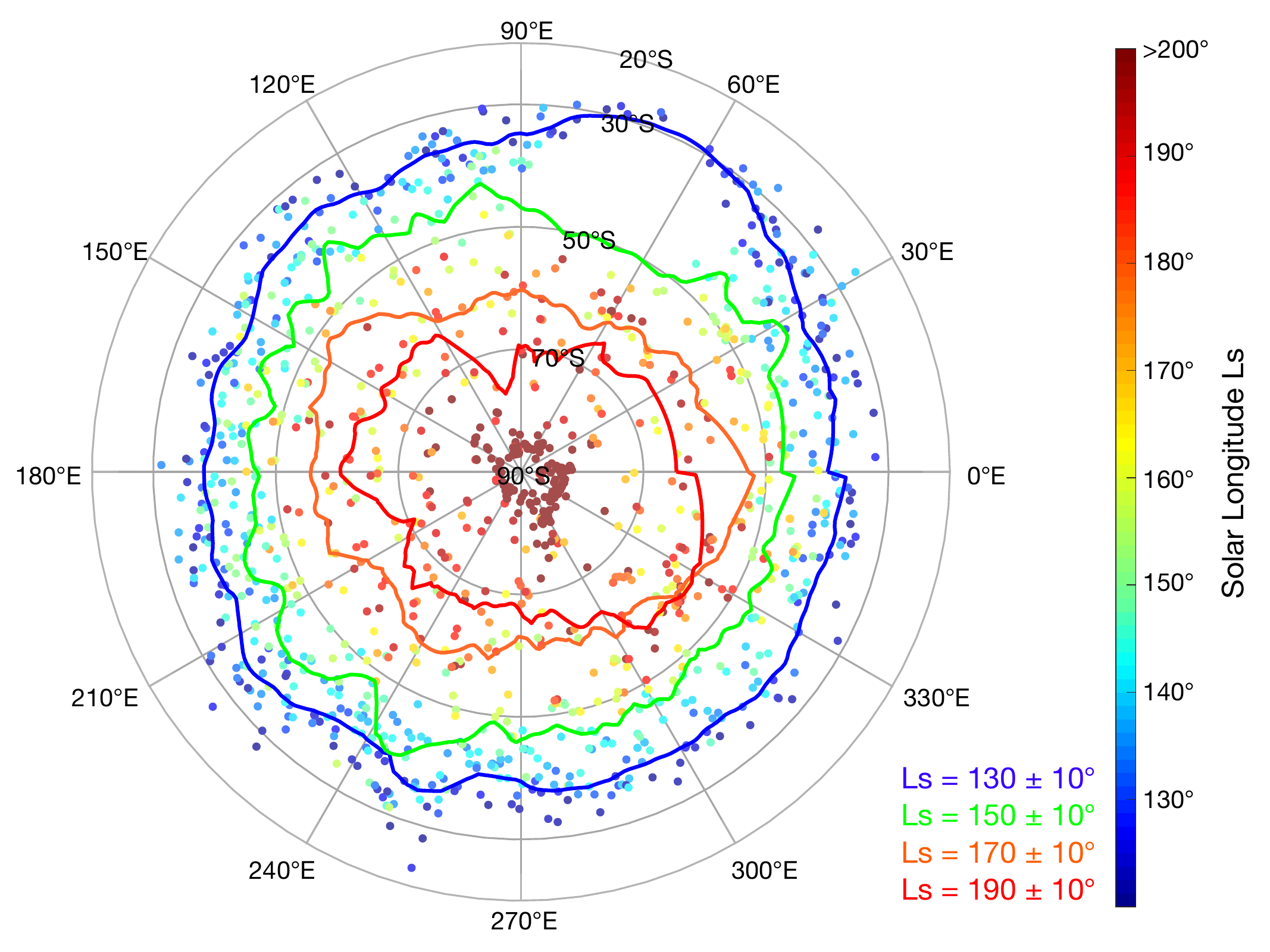}
 \caption{Spatial distribution of discontinuous \HHO~frost observed by THEMIS on flat surfaces or slopes in the Southern hemisphere (colored dots). Colors indicate the solar longitude \Ls~at the time of the acquisition. Plain curves represent median latitudinal distribution of \HHO~for \Ls~=~130\textdegree~$\pm$~10\textdegree~(blue curve), \Ls~=~150\textdegree~$\pm$~10\textdegree~(green curve), \Ls~=~170\textdegree~$\pm$~10\textdegree~(orange curve);  \Ls~=~190\textdegree~$\pm$~10\textdegree~(red curve). }
 \label{fig:recession_south}
\end{figure}

The seasonal evolution of the water frost boundary in the Southern Hemisphere is presented in Figure \ref{fig:recession_south}. At \Ls$\sim$130\textdegree~the longitudinal extent of water frost is almost homogeneous (mean latitude of 38\textdegree S~$\pm$~0.9\textdegree~at~\sigO), except in the west of Hellas which promotes the accumulation of frost as explained in section \ref{sssec:distributionspatial}. At \Ls$\sim$150\textdegree~water frost extends to 45\textdegree S~$\pm$~1.9\textdegree~at~\sigO.  In the Hellas area, water frost is not present as a consequence of the dryer air in this region (see section \ref{sssec:distributionspatial}) and observational bias.  At \Ls$\sim$170\textdegree~water frost extends to 58\textdegree S~$\pm$~2.9\textdegree~at~\sigO, except in Hellas where no frost is observed again.  Finally, at \Ls$\sim$190\textdegree,  water extends to 65\textdegree S~$\pm$~2.9\textdegree~at~\sigO. The high variability in the latitudinal extent for \Ls~=~170\textdegree~and~190\textdegree~is mostly due to the small number of detections for these \Ls~(twice less than the number of detections at Ls$\sim$130\textdegree, 150\textdegree). For all solar longitudes, the latitudinal variability also results from the variability in surface properties (albedo, thermal inertia) and where frost is detected (flat surface, steep/small slope, etc.).

For both \Ls$\sim$170\textdegree~and \Ls$\sim$190\textdegree, the latitudinal extent of the \HHO~frosts matches the extent of the seasonal \CO~ice cap \cite{Piqueux2015}. The presence of \HHO~ice deposits close to the \CO~ice cap edge can be explained by either the presence of small impurities of \HHO~ice within the \CO~ice cap or by a water ice annulus, which persists after the sublimation of the seasonal \CO~ice cap, as in the Northern Hemisphere  \cite<e.g.,>{Kieffer2001, Bibring2006, Wagstaff2008, Appr2011}. Such a possibility is discussed in section \ref{ssec:annulus}.

\subsection{Frost Temperature}
\label{ssec:frosttemp}
\begin{figure}[hbt!]
 \centering
 \includegraphics[width = 1\textwidth]{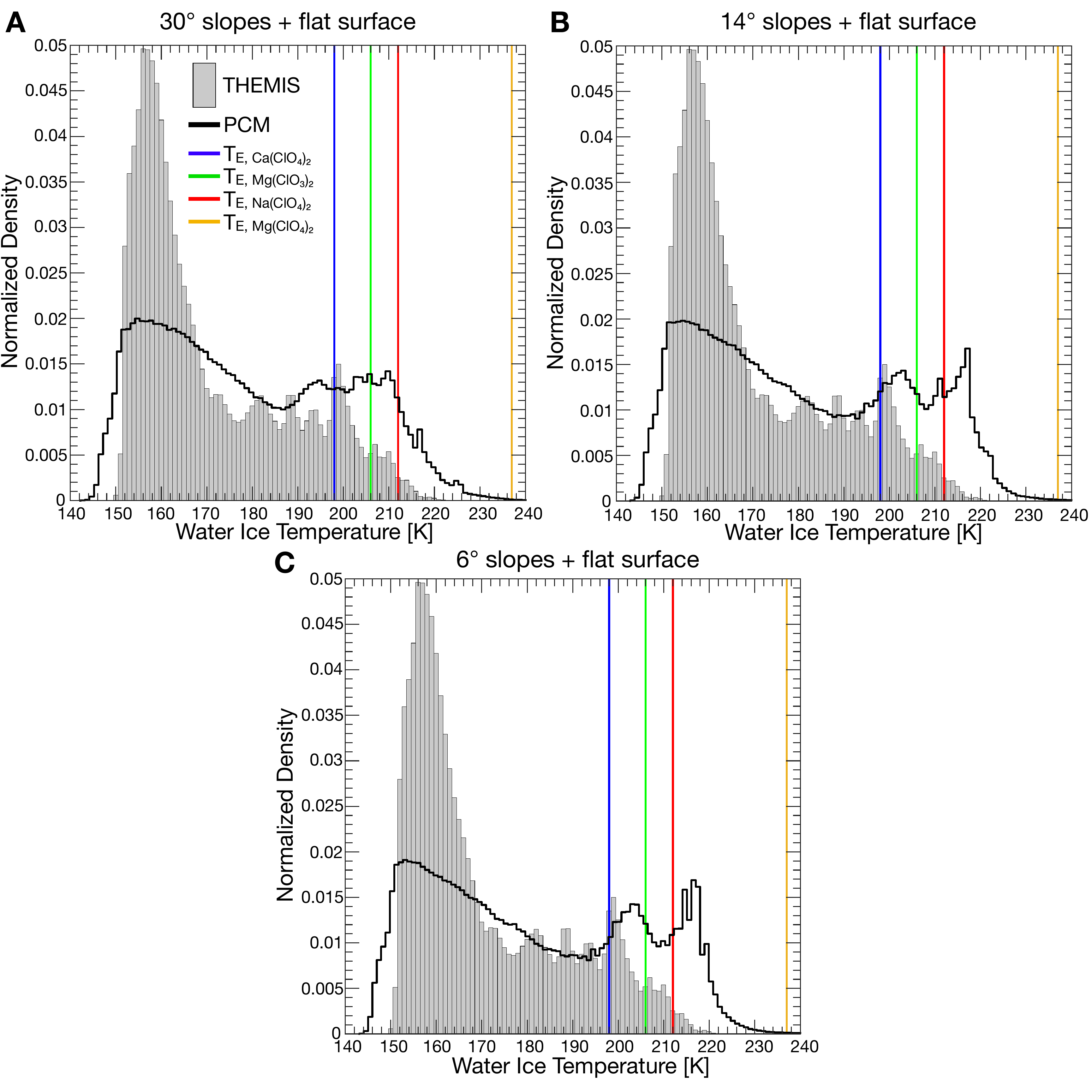}
 \caption{a) Distribution of water frost temperature measured by THEMIS (histogram) and predicted by the PCM for a flat surface, and 30\textdegree~North-facing / South-facing slopes (black curve). The distribution has been normalized by dividing each count by the total number of pixels (5.3 $\times$10$^7$ pixels for THEMIS, 3.5 $\times$10$^6$ for the PCM). The eutectic temperature of Ca-perchlorate T$_{\rm{E,~Ca(ClO}_4\rm{)}_2}$~=~198~K is represented  by the blue curve, the green curve is for the Mg-chlorate (T$_{\rm{E,~Mg(ClO}_3\rm{)}_2}$~=~206~K), the red curve is for the Mg-perchlorate (T$_{\rm{E,~Mg(ClO}_4\rm{)}_2}$~=~212~K), and the orange curve is for the Na-perchlorate (T$_{\rm{E,~Na(ClO}_4\rm{)}_2}$~=~237~K). b) Same as a), but modeled temperatures are for a flat surface, and 14\textdegree~North-facing / South-facing slopes. c) , but modeled temperatures are for a flat surface, and 6\textdegree~North-facing / South-facing slopes. } 
 \label{fig:hist_watertemp}
\end{figure}

The distribution of temperature of water ice measured by THEMIS is presented in Figure \ref{fig:hist_watertemp}. The mean temperature of water ice measured is 170.9~$\pm$~17~K~at~\sigO, with a maximum value of  253.3~K. The two peaks at 160~K and 200~K in the distribution of temperatures measured by THEMIS reflect the difference in the nature of the ices observed: The cold peak corresponds to seasonal frosts observed in the morning, after the disappearance of the \CO~ice, and the hotter peak to warm frosts observed at perennial deposit locations in summer (or to some seasonal frosts observed just before they disappear, at the end of the day). CO$_2$ frost, misinterpreted as water frost at the level of seasonal ice caps (section  \ref{ssec:annulus} ) also contributes (but to a minor extent) to the cold peak at 160~K.

In comparison, \citeA{Wagstaff2008} and \citeA{BAPST2015} found temperatures between 165~K and 210~K for the water ice deposits in the Northern Hemisphere and up to 240~K for the Southern Hemisphere \cite{BAPST2015}. Water ice temperatures measured by Carrozzo et al. (2009) on pole-facing slopes at tropical latitudes range from 180~K to 260~K. The low ice temperature measured in this study is mostly an effect of the local time of the measurements:  75\% of the water detections are made between 5 and 9 a.m. and 15\% between 5 and 9 p.m. In comparison, most of the detections made by \citeA{Carrozzo2009} are between 10 a.m. and 5 p.m. The last detections made during the rest of the day are confined to the South Pole. To quantify the bias induced by the local time, we have compared the measured temperature with the distribution of temperature predicted by the PCM for a complete MY (Figure \ref{fig:hist_watertemp}). The water ice thermal properties in the model are a broadband albedo of 0.33, thermal inertia of 800~\tiu, and an emissivity of 1 \cite{Lange2023Model}. For this computation, we extracted the temperature predicted in the PCM for a 30\textdegree~pole-facing slope, equatorward-facing slope, and flat surface only if the model predicts a frost thickness higher than 1~\mum~for these terrains. The PCM predicts water ice temperatures that are on average 180~$\pm$~21~K~at~\sigO, with a maximum temperature of 264.3~K. Hence, frost temperature measured by THEMIS seems to be lower by $\sim$10~K compared to the model. In both cases, no melting of pure water ice is expected as temperatures are below the triple point of water (273.15~K). A complete discussion on a possible melting is given in section \ref{ssec:latentheat}.

\begin{figure}[hbt!]
 \centering
     \includegraphics[scale = .3]{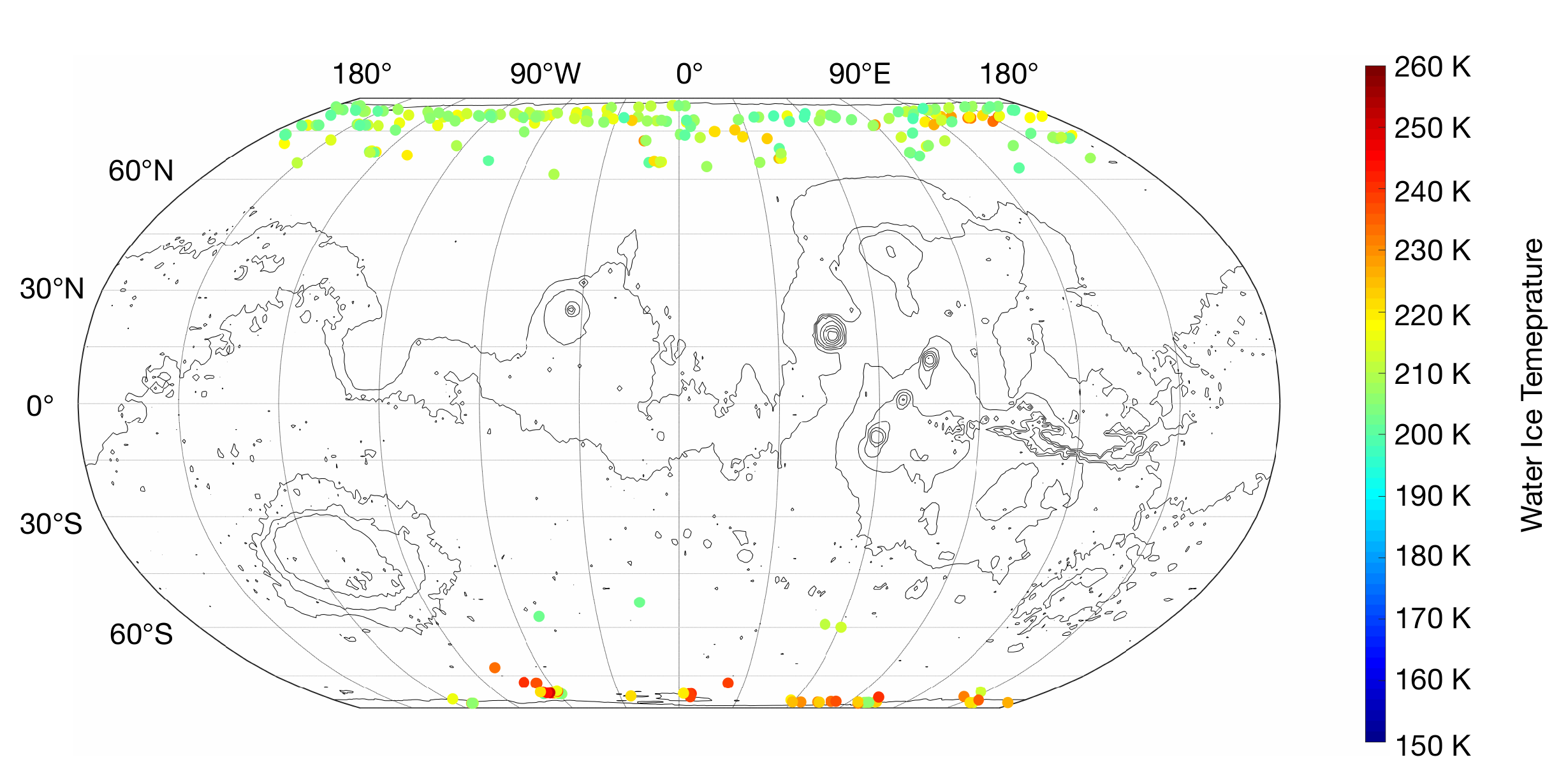}
 \caption{Distribution of the warm water frost (Temperature above T$_{\rm{E,~Ca(ClO}_4\rm{)}_2}$~=~198~K) detected with THEMIS on a Robinson projection map. Background contour lines are from MOLA topography \cite{Zuber1992}.}
 \label{fig:repartion_hotfrost}
\end{figure}

While pure water ice melting is incompatible with our results, a brine solution could form in the thermal conditions measured by THEMIS. Indeed, the presence of salts in the water ice mixture can reduce the temperature needed to melt to T$_{\rm{E,~Ca(ClO}_4\rm{)}_2}$~=~198~K for Ca-perchlorate, T$_{\rm{E,~Mg(ClO}_3\rm{)}_2}$~=~206~K for Mg-chlorate, T$_{\rm{E,~Mg(ClO}_4\rm{)}_2}$~=~212~K for Mg-perchlorate, and T$_{\rm{E,~Na(ClO}_4\rm{)}_2}$~=~237~K for Na-perchlorate \cite{Chevrier2022}. As the widespread presence of salts and perchlorates on the surface of Mars has been demonstrated \cite{Clark1981, Osterloo2008, Hecht2009, Osterloo2010, Glavin2013}, we can assume that these salts are present where we detect water ice. As shown by Figure \ref{fig:hist_watertemp}, some water ice deposits have a temperature higher than eutectic temperatures of Ca-perchlorate (267 detections, e.g. Figures \ref{fig:frost_identification}b, d), Mg-(per)chlorate (respectively 168 and 106 detections), and Na-perchlorate but to a less extent (7 detections). Their location, along with the maximum temperature of these ice deposits, are shown in Figure \ref{fig:repartion_hotfrost}.

Except for 4 images (e.g., Figure \ref{fig:frost_identification}d), all warm frosts are detected above 60\textdegree~latitude during summer  (e.g., Figure \ref{fig:frost_identification}b). For each of these sites, we have investigated THEMIS, High-Resolution Imaging Science Experiment \cite<HiRISE,>{McEwen2007} and Context Camera \cite<CTX,>{Malin2007} data acquired during summer to search for possible indications of fluid flow where frost would have melted into brines. Yet, we do not find evidence of such flow where THEMIS identified warm frosts, even where \citeA{Kereszturi2010} and \citeA{Mohlmann2010} have identified flow-like phenomena on dark dune spots at high latitudes that they attributed to brines. Some dark streak features are observed on some crater slopes (e.g., South, Schmidt craters) but can be explained by other environmental factors such as wind for instance.  We speculate that the lack of evidence of brine flow may be due to 1) the actual absence of brines and 2) if a brine does form, the amount of liquid may not be large enough to destabilize the surface material (either through lack of salt or immediate evaporation of the liquid). As water frosts are expected to be thin (from tens of micrometers to a few mm), it is hard to imagine that the brines would form a large flow, but should instead form a few droplets at the surface, which would percolate through the porous regolith.  As shown by \citeA{Schorghofer2020},  the warmest frosts should be located poleward of boulders. Indeed, frost on pole-facing slopes should warm slowly during early spring, while areas behind boulders should warm suddenly during late spring / early summer after being shadowed during part of the year. Yet, THEMIS resolution is not sufficient to observe these locations.

If these ices were to form a brine, their stability would be very limited. As demonstrated by \citeA{RiveraValentn2020} and \citeA{Chevrier2022}, brines would be stable for only a few hours a year at mid and high latitudes. Although their model does not consider the thermal effect of slopes, PCM simulations of the slope microclimates show that the temperature of water ice exceeds the eutectic temperature  T$_{\rm{E}}$  for only a few hours a day and on only a few sols (ranging from 1 to 10 sols). It should be noted that our criterion for stability is solely based on surface temperature, although it also depends on the water activity. A complete study of the stability of brines coupling a complete thermodynamic model and our GCM simulating slope microclimates is left for future work.

\subsection{Near-surface Water Vapor Content}
\label{ssec:nearsurfacewatervapor}

\begin{figure}[hbt!]
 \centering
 \includegraphics[scale = .35]{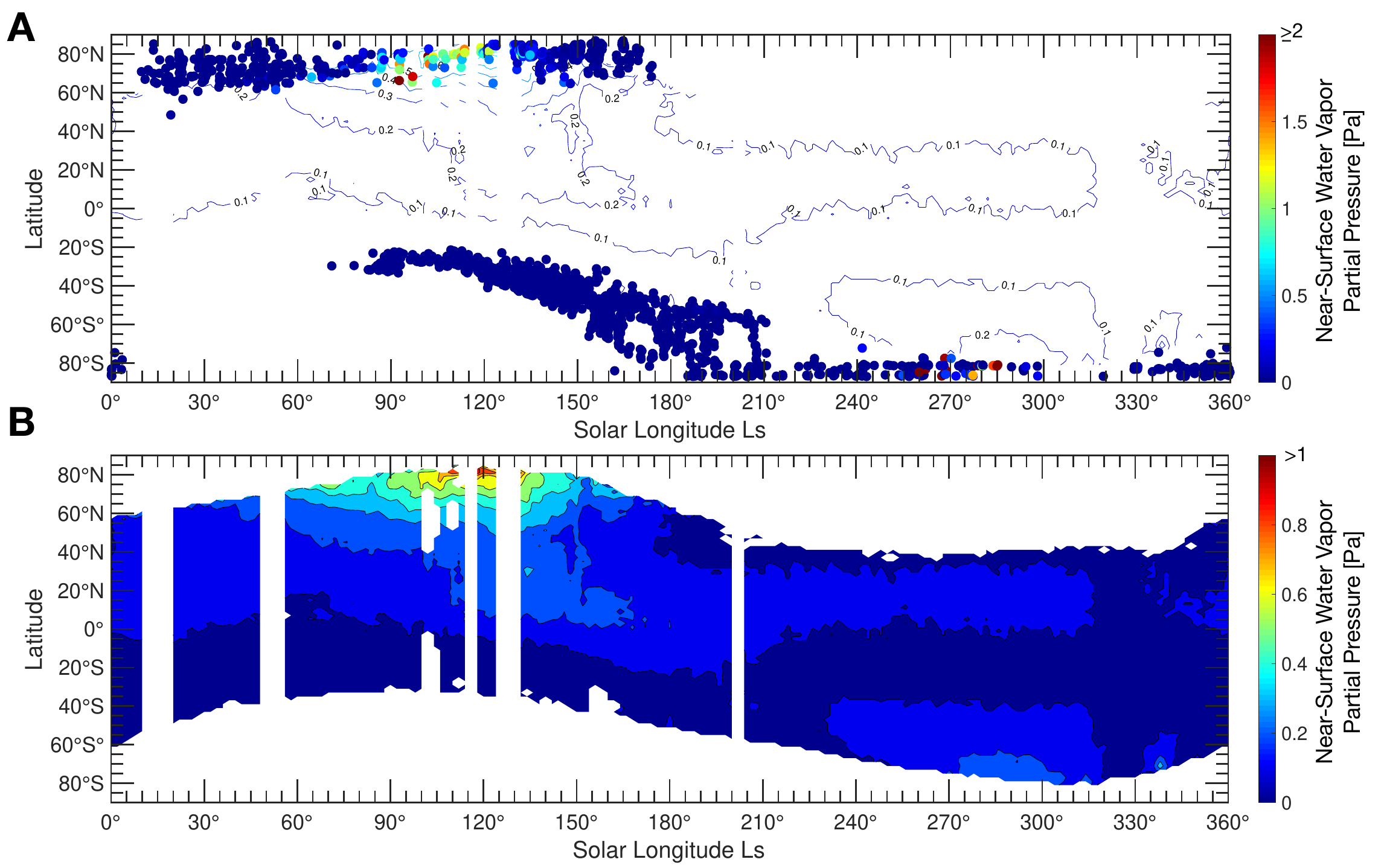}
 \caption{a) Water vapor pressure derived from ice temperature measurements with THEMIS. Background colored contours are near-surface water vapor derived from TES measurements presented in panel b). The color of the contours in a) follows the color bar from panel b).  }
 \label{fig:compTHEMISTES}
\end{figure}

Here we present the near-surface water vapor inferred from temperatures measured with THEMIS. Once $T_{\rm{ice}}$ has been measured, we derive the near-surface water vapor with Eq. \ref{eq:murphypsv}. Results are presented in Figure \ref{fig:compTHEMISTES}a. The mean partial pressure derived is 0.05$^{+0.2}_{-0.05}$~Pa~at~\sigO. The highest partial pressures are found at the poles during summer when perennial water ice caps sublime. Otherwise, most of our near-surface water vapor values are low, mostly because measurements are made during the early stage of the ice sublimation, early in the morning or late afternoon, when ice temperatures are low. 

We have nevertheless checked the consistency of our results with TES measurements of water vapor column-abundance  \cite{Smith2002}. These measurements first need to be interpolated from column-abundance to near-surface. However, such an operation is not easy as the vertical structure of the water vapor in the lowest layer of the atmosphere is not very well constrained \cite{Tamppari2020, Leung2024}. We used here the data from \citeA{Khuller2021frost}) who derived near-surface water vapor content from TES data assuming a well-mixed, hydrostatic, and isothermal atmosphere up to the H$_2$O condensation level \cite<as done in>{Schorghofer2005}. Values obtained are presented in Figure \ref{fig:compTHEMISTES}b. With this approach, the average surface water vapor for the Northern Hemisphere is 0.17~Pa and 0.09~Pa for the South \cite{Schorghofer2005}.  In comparison, the mean water vapor pressure derived from THEMIS temperature measurements is 0.12~Pa in the North and 0.0057~Pa in the South. Again, these measurements might be lower because of their time of acquisition (early morning / late afternoon) in comparison with the time of TES measurements (2 p.m., when the atmospheric water content is at maximum). Yet, the seasonal evolution of water vapor between TES and THEMIS is consistent: The maximum amount of water vapor is at the poles during the summer, and the atmosphere surrounding the seasonal CO$_2$ ice cap is dry. For very specific locations (e.g., above 70\textdegree N, 80\textdegree~$\le$~\Ls~$\le$~110\textdegree, THEMIS measurements show that the near-surface is enriched with water vapor. Indeed, at this time of the year, the sublimation of massive water ice deposits at the surface is supplying the dry atmosphere with water vapor. However, the sparse number of such defections prevents any generalization on a near-surface enrichment of water vapor. 

\begin{figure}[hbt!]
 \centering
 \includegraphics[scale = .35]{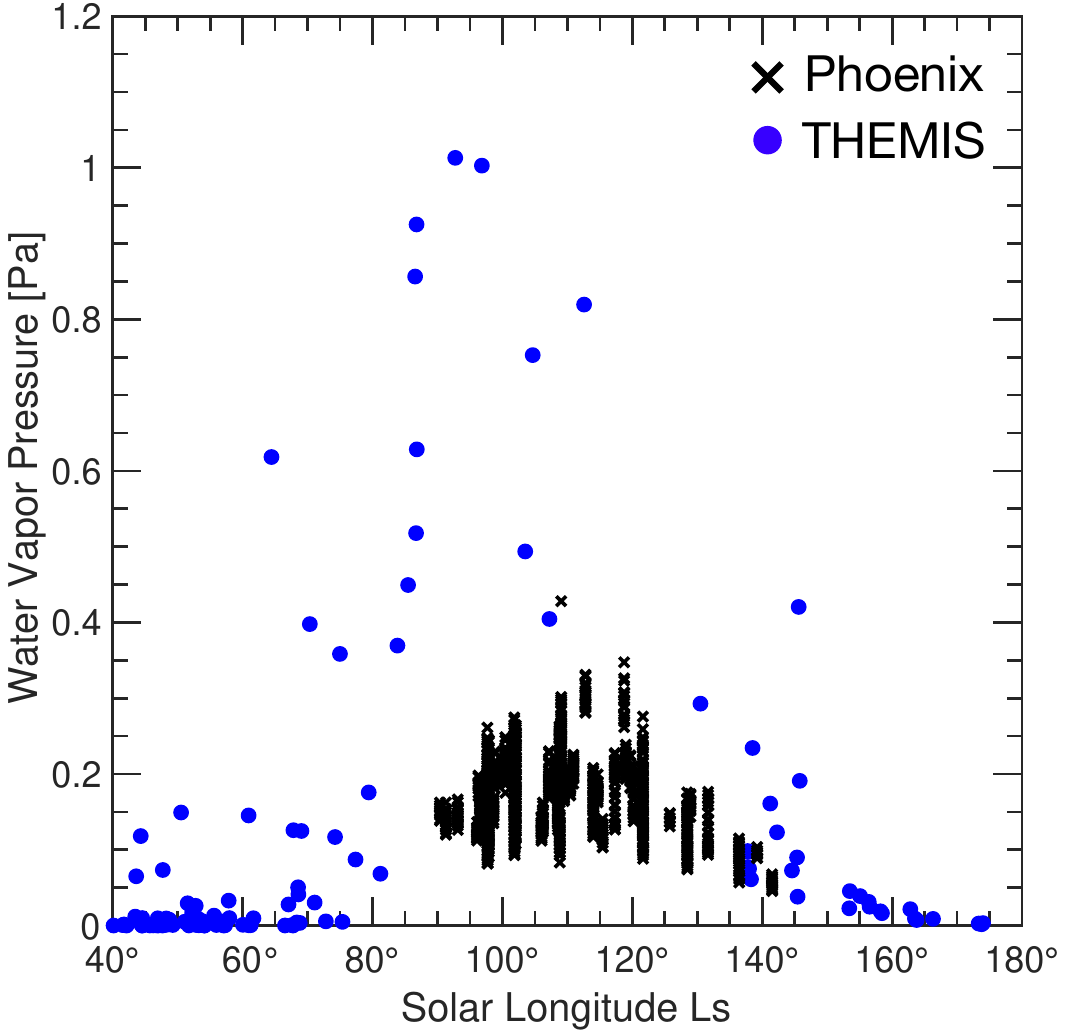}
 \caption{Water vapor pressure measured by Phoenix, using the calibration from \citeA{Fischer2019} at 6–8 a.m. and 6–8 p.m. (dark crosses) and water vapor pressure derived from ice temperatures measured with THEMIS between 65\textdegree N and 75\textdegree N (blue points). }
 \label{fig:compTHEMISPhoenix}
\end{figure}

We also present for completeness the comparison between THEMIS and Phoenix humidity measurements \cite{Fischer2019} in Figure \ref{fig:compTHEMISPhoenix}. As THEMIS does not specifically cover the Phoenix site (68.2\textdegree N, 234.3\textdegree E), we compare their measurements with those taken by THEMIS between 65\textdegree N and 75\textdegree N, without any longitude filter, assuming that it is representative of the Phoenix site. As THEMIS acquisitions were made between 6 a.m. and 8 a.m. and 6 p.m. and 8 p.m., we have isolated these data in the Phoenix dataset. Both datasets mostly overlap during the second half of the Phoenix mission, between \Ls~=~125\textdegree~and \Ls~=~145\textdegree. In both cases, water vapor measurements are consistent, with values between 0.1 and 0.3~Pa, suggesting that our vapor pressure measurements are reliable. However, observational limitations (not always the same site observed, measurements procured at particular local times only when frost is present), prevent us from extending the intercomparison.

\section{Discussion \label{sec:discussion}}

\subsection{Perennial vs. Seasonal vs. Diurnal water ice}
\label{ssec:perenvsdiurnvsseason}

Water ice deposits detected in this study could be either perennial, seasonal, or diurnal ice. Frosts detected at high latitudes during summer in the North are always found where massive ice deposits are observed and are thus considered perennial deposits (e.g., Figures \ref{fig:frost_identification}a, b). At lower latitudes, frosts detected with THEMIS can be either diurnal or seasonal.

During the night, as the surface cools, a thin layer of water frost can form, which then sublimes during the morning. Phoenix's in-situ imagery revealed such diurnal frost (estimated to be a few micrometers thick, according to our model), even though the ice signature on the surface was very weak \cite<see Figure 3b of>{Smith2009}. \citeA{Svitek1990} have also detected diurnal water frost during spring at the Viking 2 landing site, noting that the brighter areas were linked to the thickest frosts. Finally, \citeA{Landis2007} detected frost at the Opportunity rover landing sites, but it was located on the deck of the rover and not on the surface. On the other hand, at the Curiosity landing site (4.5\textdegree S), diurnal water frost was assumed to be present given the environmental conditions of temperature and water vapor \cite{MARTINEZ2016} but was not formally detected with cameras and spectrometers \cite{Schroder2015}.

\begin{figure}[hbt!]
 \centering
 \includegraphics[scale = .4]{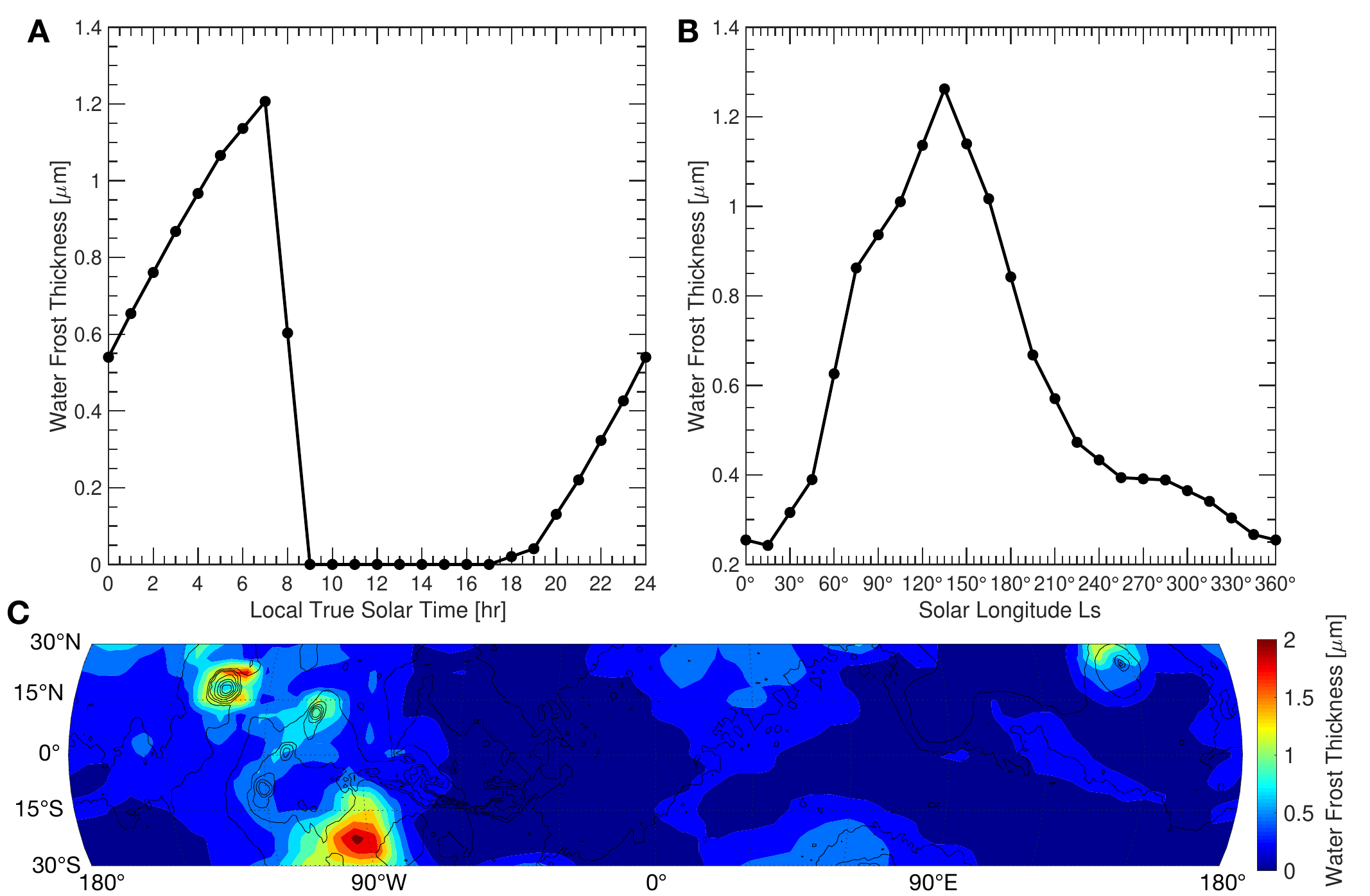}
 \caption{a) Evolution of \HHO~frost thickness predicted by the PCM at 15\textdegree N, 225\textdegree E, at \Ls~=~120\textdegree~through the day. Dots represent the outputs from the PCM, which have been linearly interpolated (plain curve). b) Evolution of \HHO~frost thickness predicted by the PCM at 6~a.m., at 15\textdegree N, 225\textdegree E. c) Map of the frost thickness predicted by the PCM at 6~a.m., \Ls~=~120\textdegree~between 30\textdegree S and 30\textdegree N on a Robinson projection map. Background contour lines are from MOLA topography \cite{Zuber1992}. For the computations of the frost thicknesses, the local slope in the PCM grid (1\textdegree~of resolution in latitude-longitude) is computed with MOLA data.} 
 \label{fig:diurnalfrost}
\end{figure}

The Mars PCM predicts the formation of this diurnal frost during the night, quickly disappearing in the early morning (Figure \ref{fig:diurnalfrost}a) during most of the year. For mid-to-high latitudes (above $\pm$45\textdegree~latitude), the diurnal thickness of water frost is at a maximum $\sim2-10$~\mum~during the early spring and late autumn while being 1–2~\mum~thick during summer. At low latitudes ($\pm$30\textdegree~latitude), frost thickness is at a maximum during northern summer, when the atmosphere is enriched in H$_2$O due to the sublimation of the northern perennial water ice cap \cite{Smith2002} but is no thicker than 1-2~\mum~(Figure \ref{fig:diurnalfrost}b). The diurnal water frost thickness during northern summer (\Ls~=~120\textdegree) predicted by the PCM between latitude 30\textdegree N and 30\textdegree S is shown in Figure \ref{fig:diurnalfrost}c. In comparison, diurnal CO$_2$ frost is expected to be 10–100~\mum~thick, i.e., ten times larger than water ice \cite{Piqueux2016}. We acknowledge that the H$_2$O frost thicknesses computed by the PCM are upper limit since the model does not account for adsorption/desorption and exchange with the regolith. As shown by  \citeA{Jakosky1997, Steele2017} and \citeA{Savijrvi2018}, adsorption during the late afternoon should deplete the near-surface water vapor content, reducing the thickness of frost formed at night to less than one~\mum.

Could THEMIS detect such diurnal frost from orbit? It seems very unlikely for two reasons. First, our method mostly detects thick frosts, at least 100~\mum~thick (Figure \ref{fig:frost_thick}, section \ref{sssec:distributionspatial}), i.e., ten times more thick than what is predicted for diurnal water frost. Therefore, it is improbable that such frost produces an albedo contrast with the bare surface strong enough to be detected with our method. Second, almost all of the frosts we have detected outside the poles are located within the stability domain of seasonal frost (Figure \ref{fig:distribution_latls_frost_all}) and not outside of it, suggesting again that they are seasonal rather than diurnal. Hence, although the frosts detected in this study are mostly observed during the early morning and might at first glance be considered diurnal frosts, we conclude that they are actually seasonal frosts.

One area of interest to detect diurnal frost with THEMIS would be at the top of volcanoes, where the contrast between atmospheric and surface temperatures is strong \cite{Fan2023}, potentially creating a significant amount of frost every night (e.g., Figure \ref{fig:diurnalfrost}). We have looked at the calderas of volcanoes in the $\pm$40\textdegree~latitude band and were unable to find any clear evidence of frost deposits. Indeed, the main bluish patterns we observed did not seem to correlate with the topography and could correspond either to surface-fogs, or clouds in the atmosphere \cite<e.g., >{Inada2007, McConnochie2010}.  \citeA{Valantinas2024} were able to demonstrate the presence of diurnal water frost on the caldera floor and rim of the Tharsis volcanoes with the Colour and Stereo Surface Imaging System and High-Resolution Stereo Camera.

\subsection{On the Existence of a Water Annulus in the South }
\label{ssec:annulus}

The presence of H$_2$O ice deposits close to the CO$_2$ ice cap edge (Figures \ref{fig:distribution_latls_frost_all}, \ref{fig:recession_south}) can be explained by either small impurities of H$_2$O ice within the CO$_2$ ice cap, left as a lag when the CO$_2$ sublimes or a water ice annulus created by the sublimation-recondensation process proposed by \citeA{Houben1997}. The latter has been widely observed in the North \cite{Kieffer2001, Bibring2006, Wagstaff2008, Appr2011}. In the South, its existence is less certain.  \citeA{Titus2005} proposed the detection of an annulus at 85\textdegree S, and \citeA{BAPST2015} suggested, using TES data, the existence of an annulus in the Southern Hemisphere during spring at 45–60\textdegree S. However, \citeA{BAPST2015} acknowledged that their detection of a bright annulus might be due to the interpolation of their data or sub-grid pixel mixing and that high-resolution thermal and visible data were needed to conclude the existence of such annulus. \citeA{Langevin2007}, using OMEGA, did not observe large-scale expanses of water frost at the edge of the CO$_2$ cap but rather local signatures of water ice in the seasonal cap, considered as impurities. The few signatures of H$_2$O ice detected with OMEGA in the Southern Hemisphere at \Ls~$\le$~190\textdegree~have either been interpreted as clouds or as frost confined in the South of Hellas \cite{Langevin2007}.

\begin{figure}[hbt!]
 \centering
 \includegraphics[scale = .4]{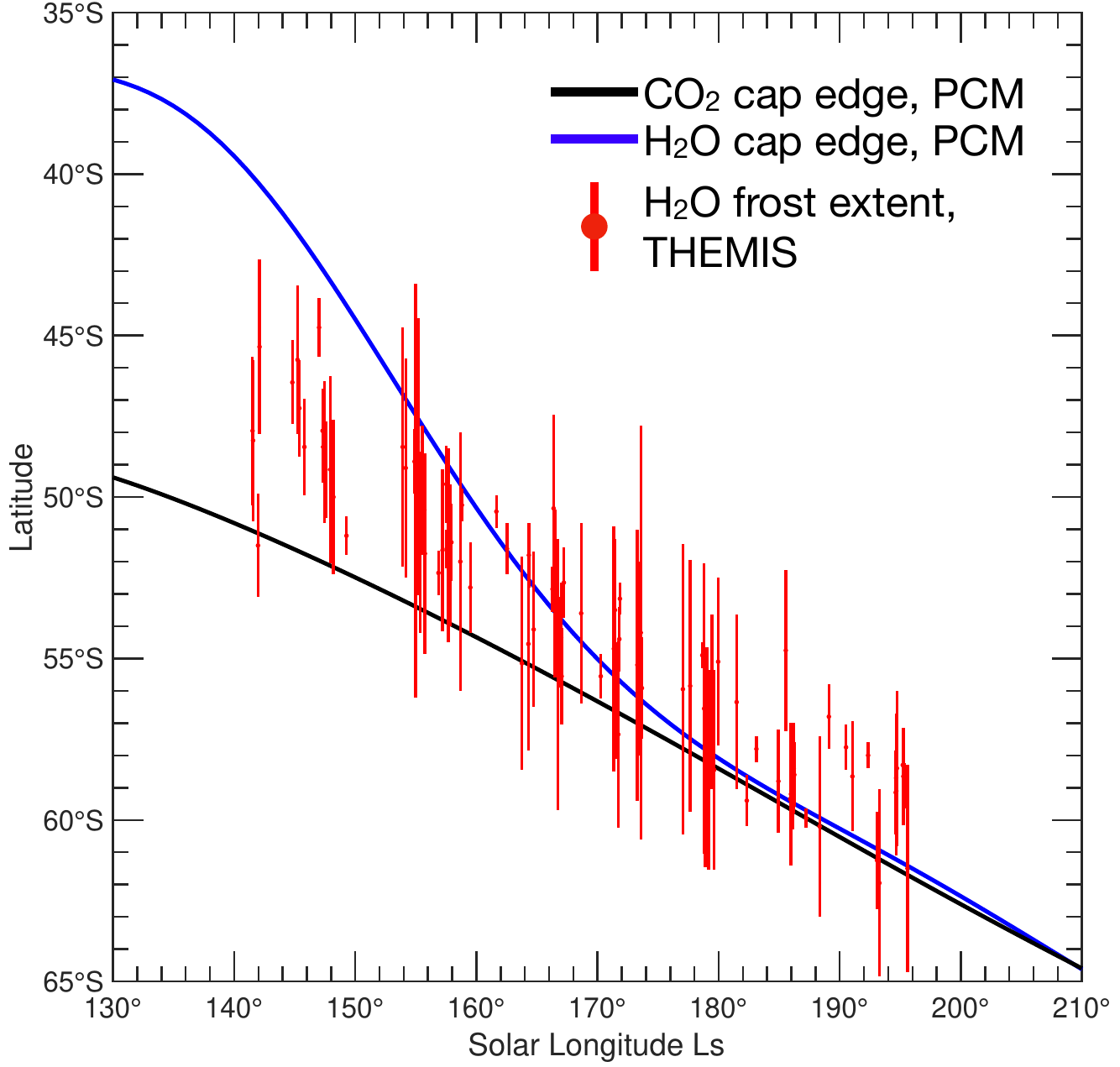}
 \caption{Latitudinal extent of H$_2$O frost deposits (red bar) during the recession of the seasonal CO$_2$ ice cap. The red bar represents the extent of H$_2$O frost, and the dots indicate their median latitudes. Plain curves represent the predicted stability of CO$_2$ (dark) and H$_2$O (blue) ice on a flat terrain by the PCM. }
 \label{fig:annulus}
\end{figure}

To test the existence of such an annulus, we have isolated the 441 detections made in the Southern Hemisphere during the receding phase of the seasonal CO$_2$ cap, i.e. at latitudes~$\le$~50\textdegree S~and 90\textdegree~$\le$~\Ls~$\le$~270\textdegree, and discriminate the nature of these deposits between H$_2$O ice impurities within the CO$_2$ cap and bright deposits close to the seasonal cap that could be interpreted as an annulus. Among these 441 water ice detections, 230 were made uniquely on pole-facing slopes and not on flat terrains, more than 1–2\textdegree~latitude from the CO$_2$ ice cap. 112 detections show isolated water ice pixels within the CO$_2$ cap, which can either be attributed to impurities or have been detected on equatorward-facing slopes where CO$_2$ ice is barely stable. Other detections are large expanses of water ice ($\ge$~0.2\textdegree~of latitudinal extent) close to the CO$_2$ cap, which could be interpreted as an annulus or frost deposits (e.g., Figures \ref{fig:frost_identification} g-h). For each of these frost deposits, we measured their extent and reported them in Figure \ref{fig:annulus}. 

Contrary to \citeA{Langevin2007}, and in agreement with \citeA{BAPST2015}, we detect water ice during the first part of the sublimation of the seasonal cap (\Ls~$\le$~190\textdegree) (e.g., Figures \ref{fig:frost_identification}c-d). These ice deposits have a small extent, on average 0.9 $^{+1.3}_{-0.9}$\textdegree~at~\sigO~of latitude, whereas the Northern annulus has an extent ranging from 5 to 10\textdegree~\cite{Kieffer2001, Bibring2006, Wagstaff2008, Appr2011}. This asymmetry is mostly explained by the asymmetry in atmospheric humidity between the two hemispheres. During winter and early spring, water ice is protected from sublimation by the cold trap formed by CO$_2$ ice. When the CO$_2$ ice disappears, the water ice is subjected to violent heating \cite{Costard2002, Schorghofer2020}, which causes it to disappear very quickly in a dry atmosphere, preventing the establishment of an extensive, stable annulus over time in the South. In comparison, in the North, the higher humidity stabilizes frost at the surface, preventing its rapid disappearance. 

Can we confidently claim the existence of an annulus in the Southern Hemisphere? Water ices detected at the edge of the cap have an average temperature of 158.8~$\pm$~4.8~K, i.e., around 10~K higher than the condensation temperature of CO$_2$. But, as noted by \citeA{Wagstaff2008}, THEMIS observations at high latitudes can lead to measured CO$_2$ temperatures of more than 170~K. As stated in section \ref{ssec:wfrostidentification_method}, we have manually eliminated these images and kept only those whose temperature is of the order of \tco~$\pm$~5~K where CO$_2$ is expected according to other observations \cite<e.g.,>{Piqueux2015}  or the PCM. However, it still appears that the thermal contrast between CO$_2$ and H$_2$O ice is not as strong as that measured by \citeA{Wagstaff2008}  (around 20 to 30 K). It is therefore possible that the low-temperature water ice is actually CO$_2$ frost. Thus, we cannot conclude whether an annulus of water ice exists in the Southern Hemisphere during spring. 

\subsection{The Effect of Latent Heat on Ice Melting/Brine Formation}
\label{ssec:latentheat}

Section \ref{ssec:frosttemp} has shown that neither ice temperatures measured by THEMIS, nor modeled by the PCM reach the melting point of 273.15~K. \citeA{Ingersoll1970}  has shown that the cooling induced by the sublimation of ice was strong enough on Mars to prevent any melting. His computations were updated by \citeA{Schorghofer2020}, who also showed that ice temperature never reached 273.15~K. In both of their models, the sublimation is only driven by the gradient of density between the CO$_2$ atmosphere and the lighter H$_2$O vapor at the surface, with an isothermal atmosphere. However, as stated in section \ref{ssec:modelh2ofrost}, the sublimation of water ice is strongly dependent on 1) near-surface atmospheric stability, 2) gustiness induced by buoyant plumes, and 3)  moisture roughness length. The state-of-the-art model by \citeA{Khuller2024}, validated in a wide range of atmospheric conditions, shows that the sublimation of water ice at $\sim$~273~K, with typical Martian atmospheric conditions, induces a significant latent heat flux, $\sim$~1000~W~m$^{-2}$, i.e., almost two times the solar energy input. Hence,  pure water ice cannot melt on present-day Mars. On the other hand, \citeA{Clow1987} and \citeA{Williams2008} have shown that taking into account the heating of dust within the water ice by solar radiation (i.e., dirty snowpack model), a solid-state greenhouse effect occurs, which can lead to ice melting.

While the release of latent heat prevents pure ice from melting, it should not prevent the melting of ice contaminated by salts.  Indeed, at the eutectic temperatures of Ca-perchlorate (T$_{\rm{E,~Ca(ClO}_4\rm{)}_2}$~=~198~K), Mg-chlorate (T$_{\rm{E,~Mg(ClO}_3\rm{)}_2}$~=~206~K), Mg-perchlorate (T$_{\rm{E,~Mg(ClO}_4\rm{)}_2}$~=~212~K), the latent heat flux associated with the sublimation of water ice is lower than 1~W~m$^{-2}$ \cite{Khuller2024}, and has a minimal impact on ice temperature. Hence, for these low temperatures, latent heat flux could not cool enough the ice to prevent the formation of brines. In this case, the limiting factor is not latent heat, but the quantity of water frost formed: as frost is only a few mm thick, it sublimates rapidly (in a few tens of sols), and can therefore disappear before it has time to reach these eutectic temperatures. However, the temperatures measured by THEMIS and modeled show that these temperatures can be reached (Figure \ref{fig:hist_watertemp}). Na-perchlorate eutectic temperatures are very unlikely to be reached (Figure \ref{fig:hist_watertemp}) as frost should sublimate before reaching this temperature, and latent heat effect should be stronger and cool the ice temperature \cite{Khuller2024}. Hence, we conclude that the more likely brine to form on Mars, based on temperature arguments, should be Ca-perchlorate, Mg-chlorate, and Mg-perchlorate rather than Na-perchlorate.

\subsection{Effect of Frost on Subsurface Ice Stability}

Massive quantities of ice have been discovered at high latitudes (above 55\textdegree~of latitude) by the Mars Odyssey Neutron Spectrometer (MONS) neutron spectrometers \cite{Boynton2002, Feldman2002, DIEZ2008}. The widely accepted theory to explain their geographical distribution and depth is that this ice is in equilibrium with near-surface water vapor \cite{Mellon1993, Mellon2004, Schorghofer2005}. Following their approach, subsurface ice is stable at a depth \textit{z}  if:

\begin{equation}
    \overline{\left(\frac{p_{\rm{vap,surf}}}{T_{\rm{surf}}}\right)} \geq \overline{\left(\frac{p_{\rm{sv}}(T_{\rm{soil}}(z))}{T_{\rm{soil}}(z)}\right)}
    \label{eq:stabilityssi}
\end{equation}

\noindent where overbars indicate time-averages over a complete MY,  $p_{\rm{vap, surf}}$~(Pa) is the vapor pressure at the surface, $T_{\rm{surf}}$~(K), is the surface temperature, $p_{\rm{sv, soil}} $~(Pa) is the saturation vapor pressure that is a function of the soil temperature $T_{\rm{soil}}$~(K) \cite{Murphy2005}. During the night and during winter, the surface is saturated. To account for this effect, \citeA{Mellon2004} and \citeA{Schorghofer2005} compute $p_{\rm{vap, surf}}$ as:

\begin{equation}
    p_{\rm{vap, surf}} = \min(p_{\rm{vap, near-surface}},p_{\rm{sv, surf}}(T_{\rm{surf}})  )
    \label{eq:effectfrost_schorgho}
\end{equation}

\noindent where $p_{\rm{vap, near-surface}}$~(Pa) is the near-surface water vapor content in the atmosphere, derived from observations or models.

\begin{figure}[hbt!]
 \centering
 \includegraphics[scale = .4]{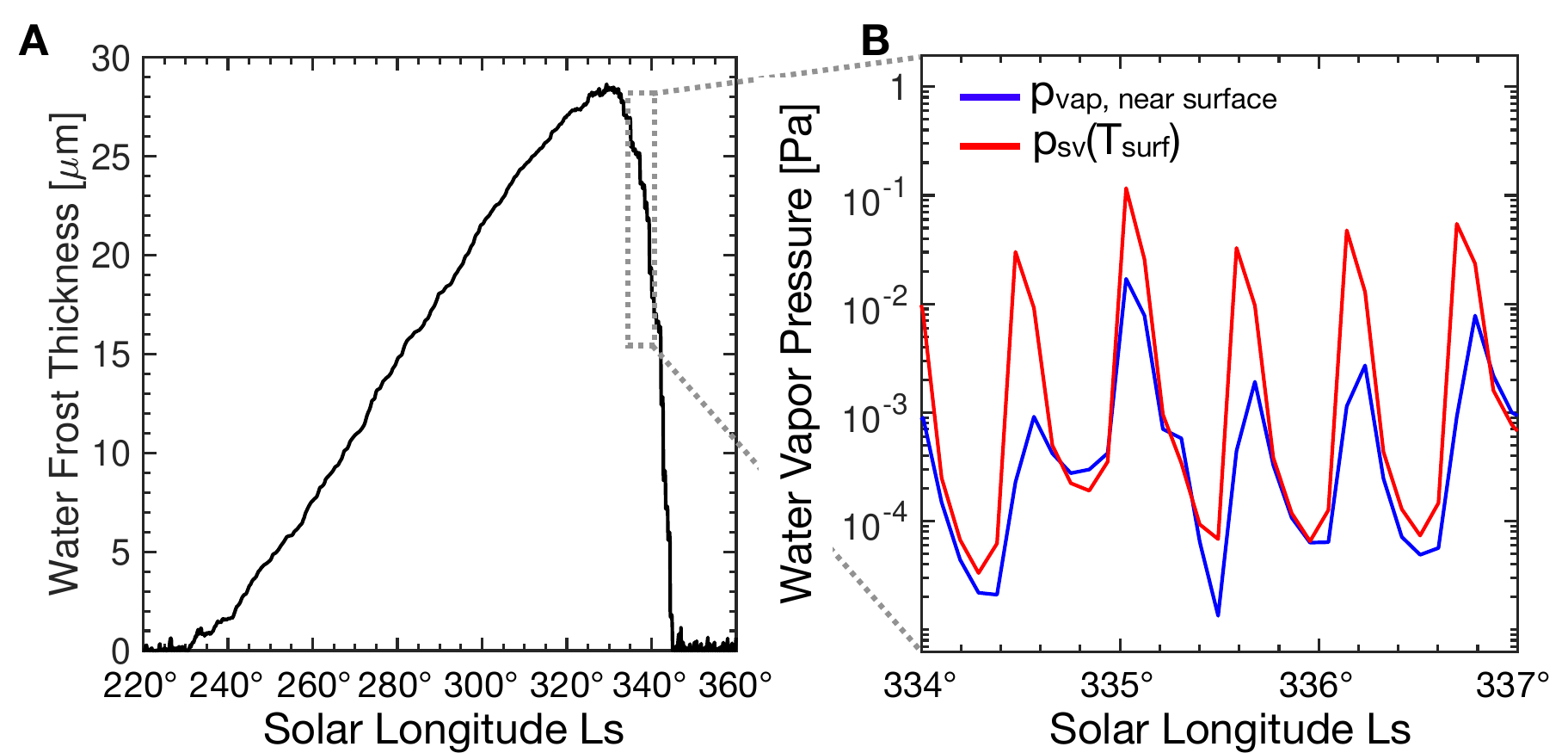}
 \caption{ a) Evolution of H$_2$O frost thickness predicted by the PCM at 52.5\textdegree N , 0\textdegree E (corresponding to a grid point where the PCM does not predict stable subsurface ice  \cite<Figure 3 of>{Lange2023ice}, while seasonal temperature variations show its presence in the first meter of the ground \cite{Piqueux2019}) during the year. b) Water vapor pressure predicted by the PCM (blue curve) in the first vertical layer ($\sim$4~m) and water vapor pressure at saturation over ice computed with Eq. \ref{eq:murphypsv} (red curve) between \Ls~=~334\textdegree~and 337\textdegree~(grey box in panel a) for the same location.}
 \label{fig:frost_ssi}
\end{figure}

However, the discovery of mid-latitude ice via ice-excavating impact craters \cite{Byrne2009, Dundas2021ice, Dundas2023} has shown that the water vapor exchange theory does not explain the presence of this ice so low in latitude with the current atmospheric humidity. Models need to double/triple the humidity to explain the stability of subsurface ice at such low latitudes \cite{Mellon2004, Lange2023ice}, suggesting that models are either underestimating near-surface humidity or that this ice is currently unstable, very slowly sublimating towards the equilibrium depth. The same problem arises for permafrost in Antarctica's dry valleys, where models predict a too-high sublimation rate of subsurface ice compared to observations \cite<see a review in >{Fisher2016}. \citeA{Hagedorn2007} and \citeA{McKay2008} have shown that snow/frost on the surface could stabilize the subsurface ice by reducing the humidity gradient between the surface and the ice table, inhibiting the sublimation loss from subsurface ice. \citeA{BAPST2015}   suggested that this effect might help solve the discrepancy in the distribution of subsurface ice between the models and the observations. In theory, the effect of surface frost is partially considered with Eq. \ref{eq:effectfrost_schorgho}. However, this equation does not consider the case when a warm frost is sublimating, and one has: $p_{\rm{sv, surf}}(T_{\rm{surf}}) \ge p_{\rm{vap, near-surface}}$, i.e. when frost sublimation replenishes the atmosphere with water vapor (e.g., section 3.3). In this case, Eq. \ref{eq:effectfrost_schorgho} underestimates the flux of water vapor coming from the surface. For an atmosphere with a water pressure of 0.2~Pa (typical of spring at mid-latitudes, Figure \ref{fig:compTHEMISTES}), this effect becomes apparent as soon as $T_{\rm{ice}}$~$\ge$~200~K, a range of temperatures measured by THEMIS (Figure \ref{fig:hist_watertemp}). GCM simulations show that this effect is mostly significant during the sublimation of seasonal/diurnal water ice (e.g., Figure \ref{fig:frost_ssi}) when warm frost acts as a significant source of water vapor in a dry atmosphere.

To quantify the effect of water frost on the yearly average near-surface water content, and therefore the stability of subsurface ice, we have changed the computation of the boundary condition given by Eq. \ref{eq:effectfrost_schorgho} with:

\begin{equation}
     p_{\rm{vap, surf}} =
        \begin{cases}
            p_{\rm{vap, near-surface}} & \mathrm{~if~frost~thickness} > 10^{-7}  \mathrm{~m} \\
p_{\rm{sv, surf}}(T_{\rm{surf}}) & \mathrm{~if~frost~thickness} < 10^{-7} \mathrm{~m}
        \end{cases}
        \label{sm}
\end{equation}
We have chosen 10$^{-7}$~m as the tolerance to avoid any effect of numerical noise. With this new boundary condition, we found that the mean annual humidity is higher by 20\% between latitude 40\textdegree N and 60\textdegree N and similarly in the South, far from the +200\% needed to model the same subsurface ice distribution as observed. In terms of depth, this means a change of mm/cm. This effect is thus not sufficient to explain the discrepancy between the ice-stability models based on Eq. \ref{eq:stabilityssi} and observations \cite{Lange2023ice}. Furthermore, as warm frosts are found locally on slopes at mid-low latitudes, the effect of warm frost on the depth of the ice table would be local, and not regional/global.

\citeA{Williams2015} have also studied the effect of surface frost on subsurface ice stability using a different approach. They have modeled the formation of subsurface ice solving the complete water vapor diffusion in the soil and found that frost has a significant impact on the depth of the stable ice table at the Viking 2 location (48\textdegree N). We cannot say whether the result comes directly from the effect of the frost on the surface or from their model. We are currently carrying out an intercomparison study between Eq.  \ref{eq:stabilityssi}–based models and full diffusion models to better understand this difference.

\section{Conclusions \label{sec:conclusions}}

The objectives of this paper are to present new detections of water ice on Mars using THEMIS images, derive their temperature, and propose a unique dataset of near-surface water vapor. The main conclusions of this investigation are:

\begin{enumerate}

\item Cross-analysis of the pixels that appear bright blue-white in visible THEMIS images with infrared temperature measurements allows us to determine whether the ice is composed of CO$_2$ or H$_2$O (Figure \ref{fig:frost_identification}). 

\item We detect water ice on 2,006 images, down to 21.4\textdegree S, 48.4\textdegree N, mostly on pole-facing slopes at low latitudes (lower than 45\textdegree~latitude) and evenly on both flat and sloped terrains at higher latitudes (Figures \ref{fig:distribution_spartial_frost}, \ref{fig:distribution_latls_frost_all}, \ref{fig:recession_south}).

\item  The evolution of water-ice deposits through the year at mid-latitude in the South is consistent with OMEGA/CRISM observations from  \citeA{Vincendon2010water} and the Mars PCM~(Figure \ref{fig:ls_omega_pcm}).

\item Like \citeA{Vincendon2010water}, and contrary to \citeA{BAPST2015}, we have been able to detect water ice in the Southern Hemisphere during autumn (Figure \ref{fig:ls_omega_pcm}). Yet, these detections are only on pole-facing slopes at low latitudes, where \citeA{BAPST2015} could not survey because of the resolution of TES. 
 
\item The mean temperature of water ice measured by THEMIS is 170.9~$\pm$~17~K~at~\sigO, with a maximum value of 253.3~K, 10~K lower than the maximum frost temperature predicted by the PCM. 267 images show water frost warm enough to enable the formation of brines (Figures \ref{fig:hist_watertemp}, \ref{fig:repartion_hotfrost}). 

\item Near-surface vapor pressures derived from ice temperatures measured by THEMIS are lower than expected based on TES data, but this difference is mostly due to the difference in acquisition local times, thus biasing the comparison  (Figure \ref{fig:compTHEMISTES}). A similar bias occurs when comparing THEMIS to Phoenix measurements (Figure \ref{fig:compTHEMISPhoenix}). 

\item Water ice frosts detected in this study are most likely related to seasonal ice rather than diurnal ice, which may be too thin to be detected (Figure \ref{fig:diurnalfrost}).

 \item We found water ice at the edge of the southern seasonal CO$_2$ ice cap during its recession. However, we cannot conclude whether there is or is not an annulus present given the small thermal contrast between water ice at the edges of the seasonal cap and the CO$_2$ ice cap itself (Figure \ref{fig:annulus}). 
 
\item  Melting of pure water ice is impossible on present-day Mars because of the cooling induced by the latent heat \cite{Ingersoll1970,Schorghofer2020,Khuller2024}. However, this conclusion does not apply to dusty ice, where a solid-state greenhouse effect might enable melting \cite{Clow1987,Williams2008}.  Brine formation is not limited by latent heat, but only by the thickness of the frost. If the frosts are too thin, they disappear before reaching eutectic temperatures. Thus, brine formation at low eutectic temperatures (Ca-perchlorate, Mg-chlorate, Mg-perchlorate) is more likely than at high eutectic temperatures (Na-perchlorate) (section \ref{ssec:latentheat}). 

\item Diurnal and seasonal water frost help to stabilize the subsurface ice by reducing the humidity gradient and acting as a source of vapor when frost sublimes. Yet, this effect is not sufficient to explain the stability of subsurface ice at mid-latitudes as revealed by impact craters excavating ice (Figure \ref{fig:frost_ssi}).

\end{enumerate}
To better constrain exchanges between the surface and the atmosphere, future work could leverage additional THEMIS bands to spectrally characterize the properties of these ices and constrain their dust content, as for instance performed with HiRISE data \cite{Khuller2021ssice}.

\section*{Open Research}
 The Mars PCM  used in this work can be downloaded with documentation at \url{https://svn.lmd.jussieu.fr/Planeto/trunk/LMDZ.MARS/.} More information and documentation are available at http://www-planets.lmd.jussieu.fr. THEMIS visible and infrared images can be retrieved from the Planetary Data System  (PDS) \cite{THEMISPDS, PDSThemisPBT}. MOLA topography map can be retrieved from the PDS \cite{Smith1999}. Data files for figures used in this analysis are available in a public repository, see \citeA{Lange2024THEMIS_data}.

\acknowledgments
This project has received funding from the European Research Council (ERC) under the European Union's Horizon 2020 research and innovation program (grant agreement No 835275, project "Mars Through Time"). Mars PCM simulations were done thanks to the High-Performance Computing (HPC) resources of Centre Informatique National de l'Enseignement Supérieur (CINES) under the allocation n\textdegree A0100110391 made by Grand Equipement National de Calcul Intensif (GENCI). Part of this work was performed at the Jet Propulsion Laboratory, California Institute of Technology under a contract with NASA (80NM0018-D0004). US Government support is acknowledged. The authors appreciate constructive comments and suggestions by D.Rogers and two anonymous reviewers which helped improve this manuscript.


%
\bibliography{agujournaltemplate.bib}
\end{document}